\documentclass[italian,english]{article}
\usepackage[T1]{fontenc}
\usepackage[latin9]{inputenc}
\usepackage{color}
\usepackage{float}
\usepackage{amsmath}
\usepackage{amssymb}
\usepackage{graphicx}
\usepackage{esint}

\makeatletter

\newcommand{\lyxmathsym}[1]{\ifmmode\begingroup\def\b@ld{bold}
  \text{\ifx\math@version\b@ld\bfseries\fi#1}\endgroup\else#1\fi}

\newcommand{\lyxaddress}[1]{
\par {\raggedright #1
\vspace{1.4em}
\noindent\par}
}

\makeatother

\usepackage{babel}
\begin{document}

\title{\textbf{f(R) gravity, relic coherent gravitons and optical chaos}}

\author{\textbf{$^{1}$ Lawrence B. Crowell and $^{2}$Christian Corda}}

\maketitle

\lyxaddress{$^{1}$Alpha Institute of Advanced Study 10600 Cibola Lp 311 NW Albuquerque,
NM 87114 also 11 Rutafa Street, H-1165 Budapest, Hungary, email: \textcolor{blue}{lcrowell@swcp.com}; }

\lyxaddress{$^{2}$Dipartimento di Fisica e Chimica, Università Santa Rita di
Firenze, Italy, International Institute for Applicable Mathematics
\& Information Sciences (IIAMIS),  Hyderabad (India) \& Udine (Italy)
and Institute for Theoretical Physics and Advanced Mathematics Einstein-Galilei
(IFM), Via Santa Gonda 14, 59100 Prato, Italy, email: \textcolor{blue}{cordac.galilei@gmail.com}.}
\begin{abstract}
We discuss the production of massive relic coherent gravitons in a
particular class of $f(R)$ gravity which arises from string theory
and their possible imprint in Cosmic Microwave Background. In fact,
in the very early universe these relic gravitons could have acted
as slow gravity waves. They may have then acted to focus the geodesics
of radiation and matter. Therefore, their imprint on the later evolution
of the universe could appear as filaments and domain wall in the Universe
today. In that case, the effect on Cosmic Microwave Background should
be analogous to the effect of water waves, which, in focusing light,
create optical caustics which are commonly seen on the bottom of swimming
pools. We analyze this important issue by showing how relic massive
GWs perturb the trajectories of Cosmic Microwave Background photons
(gravitational lensing by relic GWs).

The consequence of the type of physics discussed is outlined by illustrating
an amplification of what might be called optical chaos. 
\end{abstract}

\section{Introduction}

Modified gravity currently obtains a lot of attention from the scientific
community. The main reason is the remarkable issue that it enables
a description of early-time inflation as well as late-time acceleration
epoch (Dark Energy) in a unified way. 

In recent years, superstring/M theory caused a lot of interest about
higher order gravity in more than 4 dimensions \cite{key-1}. These
models work in the effective low-energy action of superstring theory
\cite{key-1,key-2}. Within the classical framework, they have to
be inserted among the class of the so-called $f(R)$ theories of gravity
(for a recent review see \cite{key-3}). 

Motivations for a potential extension of Einstein's general relativity
(GR) \cite{key-4} are various. First of all, as distinct from other
field theories, like the electromagnetic theory, GR is very difficult
to quantize. This fact rules out the possibility of treating gravitation
like other quantum theories and precludes the unification of gravity
with other interactions. At the present time, it is not possible to
realize a consistent quantum gravity theory which leads to the unification
of gravitation with the other forces. One of the most important goals
of modern physics is to obtain an \emph{unified theory} which could,
in principle, show the fundamental interactions as different forms
of the same \emph{symmetry}. Considering this point of view, today
one observes and tests the results of one or more breaks of symmetry.
In this way, it is possible to say that we live in an \emph{unsymmetrical}
world \cite{key-5}. In the last 60 years, the dominant idea has been
that a fundamental description of physical interactions arises from
quantum field theory \cite{key-6}. In this approach, different states
of a physical system are represented by vectors in a Hilbert space
defined in a space-time, while physical fields are represented by
operators (i.e. linear transformations) on such a Hilbert space. The
greatest problem is that this quantum mechanical framework is not
consistent with gravitation, because this particular field, i.e. the
metric $g_{\mu\nu}$, describes both the dynamical aspects of gravity
and the space-time background \cite{key-5}. In other words, one says
that the quantization of dynamical degrees of freedom of the gravitational
field is meant to give a quantum-mechanical description of the space-time.
This is an unequaled problem in the context of quantum field theories,
because the other theories are founded on a fixed space-time background,
which is treated like a classical continuum. Thus, at the present
time, an absolute quantum gravity theory, which implies a total unification
of various interactions, has not been obtained \cite{key-5}. In addition,
GR assumes a classical description of the matter which is totally
inappropriate at subatomic scales, which are the scales of the early
Universe \cite{key-3,key-5}.

In the general context of cosmological evidence, there are also other
considerations which suggest an extension of GR \cite{key-3,key-7}.
As a matter of fact, the accelerated expansion of the Universe, which
is observed today, implies that cosmological dynamics is dominated
by the so called \emph{Dark Energy}, which gives a large negative
pressure. This is the standard picture, in which this new ingredient
should be some form of un-clustered, non-zero vacuum energy which,
together with the clustered \emph{Dark Matter}, drives the global
dynamics. This is the so called \textquotedblleft{}\emph{concordance
mode}l\textquotedblright{} ($\Lambda$CDM) which gives, in agreement
with the \emph{Cosmic Microwave Background Radiation}, \emph{Large
Scale Structure} and \emph{Supernovae Ia} data, a good picture of
the observed Universe today, but presents several shortcomings such
as the well known \textquotedblleft{}\emph{coincidence}\textquotedblright{}
and \textquotedblleft{}\emph{Cosmological Constant}\textquotedblright{}
problems \cite{key-8}. 

An alternative approach is seeing if the observed cosmic dynamics
can be achieved through an extension of GR \cite{key-3,key-7}. In
this different context, it is not required to find candidates for
Dark Energy and Dark Matter, that, till now, have not been found;
only the \textquotedblleft{}\emph{observed}\textquotedblright{} ingredients,
which are curvature and baryon matter, have to be taken into account.
Then, Dark Energy and Dark Matter have to be considered like pure
effects of the presence of an intrinsic curvature in the Universe.
Considering this point of view, one can think that gravity is different
at various scales and there is room for alternative theories. 

Note that we are not claiming that GR is wrong. It is well known that,
even in the context of extended theories of gravity, GR \textit{remains
the most important part of the structure }\cite{key-7}. We are only
trying to understand if weak modifies on such a structure could be
needed to solve some theoretical and current observational problems.
In this picture, we also recall that even Einstein tried to modify
the framework of GR by adding the \textquotedblleft{}\emph{Cosmological
Constant}\textquotedblright{} \cite{key-9}. In any case, Cosmology
and Solar System tests show that modifications of GR in the sense
of extended theories of gravity have to be very weak \cite{key-3,key-7}. 

In principle, the most popular Dark Energy and Dark Matter models
can be achieved in the framework of extended theories of gravity,
i.e. $f(R)$ theories of gravity \cite{key-3} and scalar tensor theories
of gravity \cite{key-7} which are generalizations of the Jordan-Fierz-Brans-Dicke
Theory \cite{key-10,key-11,key-12}. One assumes that geometry (for
example the Ricci curvature scalar $R$) interacts with material quantum
fields generating back-reactions which modify the gravitational action
adding interaction terms (examples are high-order terms in the Ricci
scalar and/or in the Ricci tensor and non minimal coupling between
matter and gravity). This approach enables the modify of the Lagrangian,
with respect to the standard Einstein-Hilbert gravitational Lagrangian
\cite{key-13}, through the addition of high-order terms in the curvature
invariants (terms like $R^{2}$, $R^{\alpha\beta}R_{\alpha\beta}$,
$R^{\alpha\beta\gamma\delta}R_{\alpha\beta\gamma\delta}$, $R\Box R$,
$R\Box^{k}R$, in the sense of $f(R)$ Theories \cite{key-3,key-7})
and/or terms with scalar fields non-minimally coupled to geometry
(terms like $\phi^{2}R$) in the sense of Scalar-Tensor Theories \cite{key-7}. 

In the tapestry of $f(R)$ theories, the higher order terms are physically
a type of back reaction from geometry acting upon matter which further
modifies geometry. This is a topological massive gravity which represents
a form of intrinsic curvature to spacetime. These terms are related
to the Bel-Robinson tensor \cite{key-14}

\begin{equation}
{T^{\mu}}_{\nu\sigma\rho}={R^{\mu\alpha\beta}}_{\sigma}R_{\nu\alpha\beta\rho}+{R^{\mu\alpha\beta}}R_{\nu\alpha\beta\rho}-\frac{1}{2}{\delta^{\mu}}_{\nu}{R^{\alpha\beta\gamma}}_{\sigma}R_{\alpha\beta\gamma\rho}.\label{eq: Bel-Robinson}
\end{equation}
Contraction over indices gives the result

\begin{equation}
{\delta^{\sigma}}_{\mu}g^{\nu\rho}{T^{\mu}}_{\nu\sigma\rho}=R^{\mu\alpha}R_{\mu\beta}+{R^{\mu\alpha\beta}}{R^{\nu}}_{\alpha\beta\mu}-\frac{1}{2}R^{\alpha\beta\gamma\nu}R_{\alpha\beta\gamma\nu}.\label{eq: Bel-Robinson result}
\end{equation}
The physical consequences of this extension to curvature are fairly
remarkable. The Bel-Robinson tensor is a vacuum curvature $\nabla T=0$,
and it predicts gravity waves (GWs).

\section{Gravity waves in $f(R)$ theories}

In $f(R)$ gravity the GWs have longitudinal structure \cite{key-7,key-15,key-16},
which makes them comparable to acoustical waves in a media. The linearized
theory of weak GWs with a metric perturbation \cite{key-15,key-16,key-17}

\begin{equation}
g_{mu\nu}=\eta_{\mu\nu}+h_{\mu\nu}\label{eq: perturbation}
\end{equation}
gives a traceless solution in standard GR \cite{key-17}

\begin{equation}
\square{\tilde{h}}=0.\label{eq: traceless solution}
\end{equation}
The modified gravity results in a terms that acts as a mass, where
the wave equation is \cite{key-15,key-16}

\begin{equation}
\square{\tilde{h}}=m^{2}{\tilde{h}}.\label{eq: massive solution}
\end{equation}
The decomposition of the solution gives the standard $h_{++}$ and
$h_{\times\times}$ polarization modes, the mass introduces a third
polarization which is a longitudinal mode \cite{key-15,key-16}.

Let us consider a string theory setting \cite{key-1}. The gravitational
action is expanded in powers of $\alpha^{n}R^{2n}$ \cite{key-2},
for $\alpha$ the string parameter. The action is \cite{key-2}

\begin{equation}
S=\int\Big[\frac{1}{{2\kappa}}R+\alpha'R^{\mu\nu\sigma\rho}R_{\mu\nu\sigma\rho}+L\Big]\sqrt{-g}d^{4}x\label{eq: azione}
\end{equation}
being $L$ the Lagrangian for everything else compactified on a $Dp$-brane.
This action maybe trivially rewritten as standard $R^{2}$ gravity
\cite{key-44}. Following the advice \cite{key-44}, by using the
Gauss Bonnet identity (its invariant) \cite{key-45} - \cite{key-57}
one can indeed express Riemann tensor squared term as combination
of pure $R^{2}$ and Ricci tensor squared term. In such form the action
(\ref{eq: azione}) has been already studied by number of researchers,
see \cite{key-3,key-31,key-59,key-60,key-61} and references within.
This a key point. In fact, although in the form (\ref{eq: azione})
this action looks to be neither renormalizable nor ghost-free theory
\cite{key-44}, by using the Gauss Bonnet invariant one can reduce
it to the simpler form of $R^{2}$ gravity, which is the simplest
one among the class of viable models with $R^{m}$ terms in addition
to the Einstein-Hilbert theory. In Ref. \cite{key-61}, it has been
shown that such models may lead to the (cosmological constant or quintessence)
acceleration of the universe as well as an early time era of inflation.
Moreover, they seem to pass the Solar System tests, i.e. they have
the acceptable newtonian limit, no instabilities and no Brans-Dicke
problem (decoupling of the scalar) in the scalar-tensor version. 

The extremization of eq. (\ref{eq: azione}) gives

\begin{equation}
\begin{array}{c}
\delta S=0=\int\Big[\frac{1}{{2\kappa}}\frac{{\delta R}}{{\delta g^{\mu\nu}}}+2\alpha R^{\mu\nu\sigma\rho}{\frac{\delta R_{\mu\nu\sigma\rho}}{{\delta g^{\mu\nu}}}+~\frac{{\delta L}}{{\delta g^{\mu\nu}}}\Big]\sqrt{-g}d^{4}x}\\
\\
+\int\Big[\frac{1}{{2\kappa}}R+\alpha R^{\mu\nu\sigma\rho}R_{\mu\nu\sigma\rho}+L\Big]\frac{2}{{\sqrt{-g}}}\frac{{\delta(-g)}}{{\delta g^{\mu\nu}}}d^{4}x.
\end{array}\label{eq: estremizzazione}
\end{equation}
This action derives a modified form of the Einstein field equation
\begin{equation}
R_{\mu\nu}+\frac{R}{2}g_{\mu\nu}+\alpha R_{\mu\nu\sigma\rho}g^{\sigma\rho}=\kappa T_{\mu\nu}.\label{eq: einstein stringarole}
\end{equation}
Now, let us consider a metric with the form (\ref{eq: perturbation}),
which is a classical expression. The quadratic term corresponds to
quantum corrections on the order of the parameter $\alpha$. We consider
this correction as due to fields $\phi_{\nu}^{\mu}$ so that the quantum
correction to the metric is 

\begin{equation}
g_{\mu\nu}=\eta_{\mu\nu}+h_{\mu\nu}+\phi_{\mu}^{\sigma}\phi_{\nu\sigma},\label{eq: metric}
\end{equation}
where we can regard $\phi_{\mu}^{\sigma}\phi_{\nu\sigma}=\delta h_{\mu\nu}$.
These fields are physically a quantum correction to the classical
gravitational radiation $h_{\mu\nu}$. In general, these fields are
quantized fields. In a string theory framework \cite{key-1} we may
define operators of the form

\begin{equation}
\phi^{\mu\nu}=\sum_{m,n=1}^{\infty}\alpha_{m-n}^{\mu}\alpha_{n}^{\nu},\label{eq: operatori}
\end{equation}
which is a harmonic oscillator quantization condition compatible with
a string theory interpretation \cite{key-1}. The graviton fields
are given by the $n=m-n=-1$ states 

\begin{equation}
\phi_{\mu}^{\sigma}\phi_{\nu\sigma}=\alpha_{-1}^{\mu}V(x)e^{ikx}\alpha_{-1}^{\nu}V(x)e^{ikx'}\label{eq: stati}
\end{equation}
such that $\alpha_{-1}^{\mu}\alpha_{-1}^{\nu}|0\rangle=|\omega_{\mu\nu}\rangle$
constructs the elementary states.

The connection terms are computed as 

\begin{equation}
\omega_{\nu\sigma}^{\mu}=\partial_{\nu}\phi_{\rho}^{\mu}\phi_{\sigma}^{\rho},\label{eq: termini di connessione}
\end{equation}
where the field is treated as a vierbein. Now, let us compute the
curvature as

\begin{equation}
{R^{\mu}}_{nu\sigma\rho}=\partial_{\sigma}\omega_{\nu\rho}^{\mu}-\partial_{\rho}\omega_{\nu\sigma}^{\mu}+[\omega_{\sigma},\omega_{\rho}]_{\nu}^{\mu}.\label{eq: curvatura}
\end{equation}
The connection terms are on the order of the fundamental length $\alpha$,
which is small enough to ignore the second order term. The quantized
graviton field may then be written in this linearized fashion as 

\begin{equation}
\begin{array}{c}
R_{\mu\nu\sigma\rho}=\partial_{\sigma}\omega_{\nu\rho}^{\mu}-\partial_{\rho}\omega_{\nu\sigma}^{\mu}=\partial_{\sigma}(\partial_{\nu}\phi_{\gamma}^{\mu}\phi_{\rho}^{\gamma})-\partial_{\rho}(\partial_{\nu}\phi_{\gamma}^{\mu}\phi_{\sigma}^{\gamma})\\
\\
=(\partial_{\sigma}\partial_{\nu}\phi_{\gamma}^{\mu})\phi_{\rho}^{\gamma}-(\partial_{\rho}\partial_{\nu}\phi_{\gamma}^{\mu})\phi_{\sigma}^{\gamma}+\partial_{\nu}\phi_{\gamma}^{\mu}\partial_{\sigma}\phi_{\rho}^{\gamma}-\partial_{\nu}\phi_{\gamma}^{\mu}\partial_{\rho}\phi_{\sigma}^{\gamma},
\end{array}\label{eq: quantized graviton field}
\end{equation}
where the last term is zero in a linearized approximation. 

The linearized approximation occurs for long wavelength gravitons.
Assume the connection term $\omega_{\nu\sigma}^{\mu}=\partial_{\nu}\phi_{\rho}^{\mu}\phi_{\sigma}^{d}$
is eigenvalued with a wave number $k^{a}$

\begin{equation}
\omega_{\nu\sigma}^{\mu}=k_{\nu}\phi_{\rho}^{\mu}\phi_{\sigma}^{d}\label{eq: omega k}
\end{equation}
so the curvature tensor is 

\begin{equation}
\begin{array}{c}
R_{\mu\nu\sigma\rho}\simeq\partial_{\sigma}\omega_{\nu\rho}^{\mu}-\partial_{\rho}\omega_{\nu\sigma}^{\mu}=\\
\\
=(k_{\sigma}k_{\nu})\phi_{\gamma}^{\mu}\phi_{\rho}^{\gamma}-{k_{\rho}k_{\nu}}\phi_{\gamma}^{\mu}\phi_{\sigma}^{\gamma}=(k_{\sigma}k_{\nu})\delta h_{\rho}^{\mu}-{k_{\rho}k_{\nu}}\delta h_{\sigma}^{\mu}.
\end{array}\label{eq: curvature tensor}
\end{equation}
The second order term in the action is then 

\begin{equation}
R_{\mu\nu\sigma\rho}R^{\mu\nu\sigma\rho}\simeq[(k_{\sigma}k_{\nu})\delta h_{\rho}^{\mu}-(k_{\rho}k_{\nu})\delta h_{\sigma}^{\mu}][(k^{\sigma}k^{\nu})\delta h_{\sigma\nu}-(k^{\rho}k^{\nu})\delta h_{\mu\sigma}]=6k^{4}.\label{eq: second order term}
\end{equation}
The term $\alpha k^{4}$ is a quartic term in mass, where the string
coupling constant $\sim G_{N}$ has naturalized units of area. This
is an intrinsic curvature in spacetime. The string coupling constant
is about $\alpha\sim10^{-60}cm^{2}$, which is a small number. This
also guarantees the viability of the action (\ref{eq: azione}) because
the theory can pass Solar System and Cosmology tests \cite{key-7}. 

Is it possible that this mass effect should then become apparent in
the laboratory? The question is, what is the laboratory? The obvious
laboratory is the Cosmic Microwave Background (CMB). In fact, we recall
that relic gravitons should have been produced in the Inflationary
Era. This is a consequence of generals assumptions. Essentially it
derives from a mixing between basic principles of classical theories
of gravity and of quantum field theory \cite{key-18,key-19,key-20}.
The strong variations of the gravitational field in the early universe
amplify the zero-point quantum oscillations and produce relic GWs.
It is well known that the detection of relic GWs is the only way to
learn about the evolution of the very early universe, up to the bounds
of the Planck epoch and the initial singularity \cite{key-18,key-19,key-20}.
It is very important to stress the unavoidable and fundamental character
of this mechanism. The model derives from the inflationary scenario
for the early Universe \cite{key-19}, which is tuned in a good way
with the WMAP data on the CMB (in particular exponential inflation
and spectral index $\approx1$) \cite{key-21,key-22}. Inflationary
models of the early Universe were analyzed in the early and middles
1980's \cite{key-19}. These are cosmological models in which the
Universe undergoes a brief phase of a very rapid expansion in early
times. In this context the expansion could be power-law or exponential
in time. Inflationary models provide solutions to the horizon and
flatness problems \cite{key-19} and contain a mechanism which creates
perturbations in all fields \cite{key-18,key-20}. Important for our
goals is that this mechanism also provides a distinctive spectrum
of relic GWs. The GWs perturbations arise from the uncertainty principle
and the spectrum of relic GWs is generated from the adiabatically-amplified
zero-point fluctuations \cite{key-18,key-20}. 

Relic gravitons can be characterized by a dimensionless spectrum \cite{key-18,key-20}
\begin{equation}
\Omega_{gw}(f)\equiv\frac{1}{\rho_{c}}\frac{d\rho_{gw}}{d\ln f},\label{eq: spettro}
\end{equation}
where 
\begin{equation}
\rho_{c}\equiv\frac{3H_{0}^{2}}{8G}\label{eq: densita' critica}
\end{equation}
is the (actual) critical density energy, $\rho_{c}$ of the Universe,
$H_{0}$ the actual value of the Hubble expansion rate and $d\rho_{gw}$
the energy density of relic GWs in the frequency range $f$ to $f+df$.

In standard inflationary model the spectrum is flat over a wide range
of frequencies, see \cite{key-18,key-20} and figure 1. The more recent
value for the flat part of the spectrum that arises from the WMAP
data can be found in \cite{key-20},

\begin{equation}
\Omega_{gw}(f)\leq9*10^{-13}\label{eq: spettro-1}
\end{equation}

\begin{figure}[H]
\includegraphics[scale=0.9]{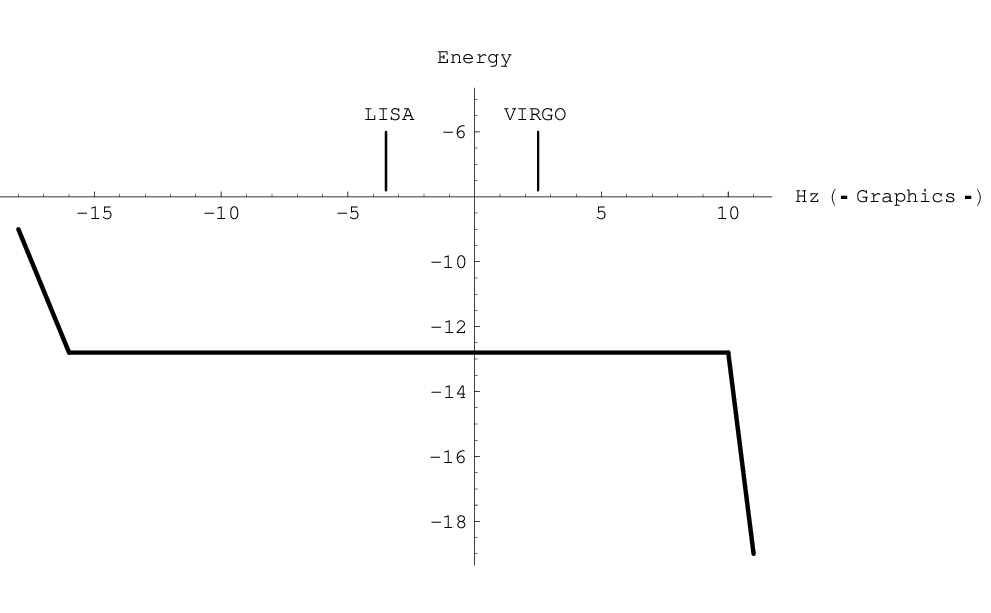}

\caption{adapted from ref. \cite{key-42} }

The spectrum of relic scalar GWs in inflationary models is flat over
a wide range of frequencies. The horizontal axis is $\log_{10}$ of
frequency, in Hz. The vertical axis is $\log_{10}\Omega_{gsw}$. The
inflationary spectrum rises quickly at low frequencies (wave which
re-entered in the Hubble sphere after the Universe became matter dominated)
and falls off above the (appropriately redshifted) frequency scale
$f_{max}$ associated with the fastest characteristic time of the
phase transition at the end of inflation. The amplitude of the flat
region depends only on the energy density during the inflationary
stage; we have chosen the largest amplitude consistent with the WMAP
constrains on scalar perturbations. This means that at LIGO and LISA
frequencies, $\Omega_{gw}(f)h_{100}^{2}<9*10^{-13}$
\end{figure}

Based on the weakness of the signal, it will be very difficult to
detect relic gravitons on Earth, but a potential detection could be,
in principle, realized with LISA \cite{key-15}. However, the presence
of relic gravitons may have perturbed the early universe in ways that
might be observable in the fine details of the CMB background. These
gravitons would introduce a small dispersion in GWs, which might then
leave an imprint on the CMB. We will discuss the potential presence
of such an imprint in next Section. 

Now, let us expand the field $\phi_{\nu}^{\mu}$ according to harmonic
oscillator operators $b,\mbox{ }b^{\dagger}$, as a simple model of
a string. The fields are expanded as

\begin{equation}
\phi_{\nu}^{\mu}=(\frac{1}{\sqrt{2}})\sum_{k}E_{\nu}^{\mu}{b(k)e^{i\theta(k)}+b^{\dagger}e^{-i\theta(k)}}\label{eq: expanded}
\end{equation}
where $E_{\nu}^{\mu}$ is a tetrad, which is discussed more below.
The summation runs from $\{-\infty,\mbox{ }\infty\}$. The product
$\phi_{a}^{c}\phi_{\nu\sigma}=\delta h_{\mu\nu}$ is a harmonic oscillator
operator

\begin{equation}
\begin{array}{c}
\phi_{\nu}^{\mu}\phi_{\sigma\mu}=(\frac{1}{2})\sum_{kk'}E_{\nu\sigma}^{2}{b(k)b^{\dagger}(k')e^{i\theta(k)-i\theta(k')}+b^{\dagger}(k)b(k')e^{-i\theta(k')-i\theta(k)}}\\
\\
+(\frac{1}{2})\sum_{kk'}E_{\nu\sigma}^{2}{b(k)b(k')e^{i\theta(k)+i\theta(k')}+b^{\dagger}(k)b^{\dagger}(k')e^{-i\theta(k)-i\theta(k')}},
\end{array}\label{eq: harmonic oscillator}
\end{equation}
The sum gives a delta function on $k$ and $k'$ and the first term
is the Hamiltonian, which after the use of a commutator the RHS term
is 

\begin{equation}
\mbox{\ensuremath{\phi_{\nu}^{\mu}\phi_{\sigma\mu}}= (\ensuremath{\frac{1}{2}})\ensuremath{\sum_{k}E_{\nu\sigma}^{2}b^{\dagger}}(k)b(k)+ (\ensuremath{\frac{1}{2}})\ensuremath{\sum_{k}E_{\nu\sigma}^{2}}{b(k)b(-k)+\ensuremath{b^{\dagger}}(k)\ensuremath{b^{\dagger}}(-k)},}\label{eq: right hand side term}
\end{equation}
where the zeta point energy (ZPE) term has been dropped. The first
RHS term is a familiar Hamiltonian type of term, while the second
term is similar to a squeeze operator in quantum optics \cite{key-23}.

The tetrad $E_{\nu}^{\mu}$ is the amplitude of the field. This plays
a role similar to the minimal electric field $E=\sqrt{\hbar\omega/V\epsilon_{0}}$
in box normalization \cite{key-24}. A plausible choice for tetrad
is then $E_{\nu}^{\mu}=\sqrt{\alpha\omega}\delta_{\nu}^{\mu}$, where
$\alpha$ is the string parameter and $\omega$ the frequency. For
$\alpha\ll1/\omega$ this is a small term. 

The curvature in quantum modes is then

\begin{equation}
R_{\mu\nu\sigma\rho}\simeq(\frac{1}{2})\sum_{k}\Big(k_{\sigma}k_{\nu}(E^{2})_{\mu}^{\beta}E_{\beta\rho}^{2}-k_{\rho}k_{\nu}E_{\mu\beta}^{2}(E^{2})_{\sigma}^{\beta})(b^{\dagger}(k)b(k)+b(k)b(-k)+b^{\dagger}(k)b^{\dagger}(-k)\Big),\label{eq: quantum modes}
\end{equation}
which is $O(\alpha)$ in the scale parameter. In fluctuations of the
curvature the metric is $g\sim\delta L/L$, for $\delta L>L_{p}$.
The connection terms are of order $\Gamma\sim\delta L/L^{2}$ and
curvatures are $R\sim\delta L/L^{3}$. The wave vectors are $k\sim1/L$
and the scaling parameter is $\alpha\omega\sim\delta L/L$. From a
dimensional and scaling perspective this answer appears at least proximal. 

For the sake of simplicity, let us write the curvature tensor as 

\begin{equation}
R_{\mu\nu\sigma\rho}\simeq(\frac{1}{2})\sum_{k}\Pi_{\mu\nu\sigma\rho}(k)\big(b^{\dagger}(k)b(k)+b(k)b(-k)+b^{\dagger}(k)b^{\dagger}(-k)\big).\label{eq: Riemnann}
\end{equation}
The second order term is formed from the total contraction on the
Riemann tensor $R_{\alpha\lyxmathsym{ß}\mu\nu}R_{\sigma\rho}^{\alpha\beta}$,
it is

\begin{equation}
\begin{array}{c}
R_{\mu\nu\sigma\rho}R^{\mu\nu\sigma\rho}\simeq(\frac{1}{4})\sum_{kk'}\Pi_{\mu\nu\sigma\rho}(k)\Pi^{\mu\nu\sigma\rho}(k')x\\
\\
\Big(b^{\dagger}(k)b(k)b^{\dagger}(k')b(k')+b^{\dagger}(k)b(k)(b(k')b(-k'+b^{\dagger}(k')b^{\dagger}(-k'))+(b(k')b(-k')+\\
\\
b^{\dagger}(k')b^{\dagger}(-k'))b^{\dagger}(k)b(k)+\big(b(k)b(-k)+b^{\dagger}(k)b^{\dagger}(-k))(b(k')b(-k')+b^{\dagger}(k')b^{\dagger}(-k')\big)\Big).
\end{array}\label{eq: sot}
\end{equation}
This term is to $O(\alpha^{2})$ and contributes a term $O(\alpha^{3})$
to the Lagrangian. 

Consider the operator matrix operation $R_{\mu\nu\sigma\rho}R^{\mu\nu\sigma\rho}|m\rangle$.
The first term has the operator matrix elements

\begin{equation}
\begin{array}{c}
b^{\dagger}(k)b(k)b^{\dagger}(k')b(k')|m\rangle=b^{\dagger}(k)\sum_{n}|n\rangle\langle n|b(k)b^{\dagger}(k')b(k')|m\rangle\\
\\
=m(k')n(k)\delta_{mn}\delta_{kk'}
\end{array}\label{eq: operator matrix elements}
\end{equation}
where $\sum_{n}|n\rangle\langle n|$ is a completeness sum and the
momentum values assumed in the states $|m\rangle$ and $|n\rangle$.
This contributes an energy-squared. A similar analysis for $\langle m|R_{\mu\nu\sigma\rho}R^{\mu\nu\sigma\rho}$
gives

\begin{equation}
\langle m|b^{\dagger}(k)b(k)(b(k')b(-k')+b^{\dagger}(k')b^{\dagger}(-k'))=m(k)(b(k')b(-k')+b^{\dagger}(k')b^{\dagger}(-k'))\label{eq: similar}
\end{equation}
and for $R_{\mu\nu\sigma\rho}R^{\mu\nu\sigma\rho}|m\rangle$ 

\begin{equation}
(b(k)b(-k)+b^{\dagger}(k)b^{\dagger}(-k))b^{\dagger}(k')b(k')|m\rangle=m(b(k)b(-k)+b^{\dagger}(k)b^{\dagger}(-k))|m\rangle.\label{eq: similar 2}
\end{equation}
The operators $b(k)b(-k)+b^{\dagger}(k)b^{\dagger}(-k)$ form the
squeeze operator \cite{key-25}

\begin{equation}
S=\exp((\frac{1}{2})(z^{*}b(k)-zb^{\dagger}(k))),\label{eq: squeeze operator}
\end{equation}
where $z^{*}=z=i((1/4)\Pi_{\mu\nu\sigma\rho}(k)\Pi^{\mu\nu\sigma\rho})$.
Hence the action phase due to the action $e^{iS}$ contains a squeeze
operator.

The final operator term is more complicated. The operator terms $b(k)b(-k)b(k)b(-k)$
and $b^{\dagger}(k)b^{\dagger}(-k)b(k)b(-k)$ are evaluated by commuting
operators and this leads to the square of number operators $n(k)n(-k)$.
The terms $b(k)b(-k)b(k)b(-k)$ and $b^{\dagger}(k)b^{\dagger}(-k)b^{\dagger}(k)b^{\dagger}(-k)$
are then a product of terms which represent a squeezed state operator. 

The squeeze operator $S(z)$ acts upon the displacement operator $D(\alpha)=\exp(\alpha b^{\dagger}-\alpha^{*}b)$
so that $S(z)D(\alpha)\ne D(\alpha)S(z)$,

\begin{equation}
\begin{array}{c}
S(z)D(\alpha)=\exp[(z^{*}b^{2}-z(b^{\dagger})^{2})/2]\exp(\alpha b^{\dagger}-\alpha^{*}b)=\\
\\
\exp[(z^{*}b^{2}-z(b^{\dagger})^{2})/2+\alpha b^{\dagger}-\alpha^{*}b)\exp[-(\frac{1}{4})(z^{*}\alpha b^{\dagger}-z\alpha^{*}b)],
\end{array}\label{eq: operator}
\end{equation}
which effectively creates a modified displacement operator 
\begin{equation}
\exp[(z^{*}b^{2}-z(b^{\dagger})^{2})/2+\alpha b^{\dagger}-\alpha^{*}b)S(z)D(\alpha)=\exp[-(\frac{1}{4})(z^{*}\alpha b^{\dagger}-z\alpha^{*}b)]=D(z^{*}\alpha).\label{eq: displacement operator}
\end{equation}
The action of the squeezed state operator on $b$ is $SbS^{\dagger}=b\cosh(|z|)+b^{\dagger}\sinh(|z|)$,
which is a Bogoliubov transformed operator \cite{key-26}. For a set
of bosons, here linear gravitons, with the same state there is then
$\sum_{n}\alpha^{n}/\sqrt{n}|n\rangle$ states with the operator acting
on this $\sum_{n}\alpha^{n}(a^{\dagger})^{n}/\sqrt{n}$ acting on
the vacuum. This operator may be formed from the $S(z)D(\alpha)S^{\dagger}(z)$
for $\alpha$ small and $|z|\gg|\alpha|$ with

\begin{equation}
S(z)D(\alpha)S^{\dagger}(z)\simeq\exp[-(\frac{1}{4})(z^{*}\alpha b^{\dagger}-z\alpha^{*}b)],\label{eq: alpha}
\end{equation}
and where we may then define $z^{*}\alpha/4\rightarrow\alpha$, and
the Bogoliubov transformation of the operator $b-b^{\dagger}$ constructs
a displacement operator \cite{key-27}.

In this way the $R^{2}$ term in the action describes the squeezed
state operator which acts on the field raising and lowering operators
to define a displacement operator for coherent states, which in the
case of photons are laser states of light. 

We then evaluate a Wilson loop \cite{key-28} $W(\phi_{\nu}^{\mu})=\exp(\int i\phi_{\mu\nu}e^{\mu}dx^{\nu}$).
In the path integral 

\begin{equation}
Z[\phi,W]=\int{\cal D}[\phi]We^{iS[\phi]}.\label{eq: path integral}
\end{equation}
The infinitesimal shift in the field $\phi\rightarrow\phi+\delta\phi$
adjusts $Z[\phi,W]\rightarrow Z[\phi+\delta\phi,W]=\langle W\rangle$
and the expansion is

\begin{equation}
\begin{array}{c}
\langle W\rangle=\int D[\phi]W(\phi+\delta\phi)e^{iS[\phi+\delta\phi]}\\
\\
=\langle W\rangle+\int D[\phi]\delta\phi\Big(\frac{{\delta W}}{{\delta\phi}}+i\frac{{W\delta S}}{{\delta\phi}}\Big)We^{iS[\phi]}),
\end{array}\label{eq: expansion}
\end{equation}
where the invariance of the expectation gives 

\begin{equation}
\frac{{\delta W}}{{\delta\phi}}+i\frac{{W\delta S}}{{\delta\phi}}=0\rightarrow\frac{{\delta ln(W)}}{{\delta\phi}}+i\frac{{\delta S}}{{\delta\phi}}=0.\label{eq: expectation}
\end{equation}
This formula is only well defined for a polynomial function. So we
make the following approximation. The loop is considered to be very
small and in that way we can approximate the Wilson loop with

\begin{equation}
W(\phi)=1+i\phi_{\mu\nu}e^{\mu}\delta x^{\nu},\label{eq: approximate Wilson loop}
\end{equation}
so that the functional derivative of $W(\phi)$ is

\begin{equation}
\frac{{\delta W}}{{\delta\phi^{\mu\nu}}}\simeq i\epsilon_{\sigma\nu}\delta_{\mu}^{\nu\sigma}\delta(x-x')\label{eq: functional derivative}
\end{equation}
for $\epsilon_{\mu}^{\nu}$ a unit area. The solution is then

\begin{equation}
\Big\langle\frac{{\delta W}}{{\delta\phi}}+iW\frac{{\delta S}}{{\delta\phi}}\Big\rangle=0\rightarrow\Big\langle W\frac{{\delta S}}{{\delta\phi^{\mu\nu}}}\Big\rangle\simeq\epsilon_{\sigma\nu}\delta_{\mu}^{\sigma}\delta(x-x').\label{eq: solution}
\end{equation}

Now, let us consider the second order expansion 

\begin{equation}
W(\phi)=1+i\phi_{\mu\nu}e^{\mu}\delta x^{\nu}+\frac{1}{2}\phi_{\mu\nu}\phi_{\sigma\rho}e^{\mu}e^{\sigma}\delta x^{\nu}\delta x^{\rho}=W^{0}(\phi)+W^{1}(\phi)+W^{2}(\phi),\label{eq: second order expansion}
\end{equation}
which gives the result

\begin{equation}
\Big\langle W^{2}\frac{{\delta S}}{{\delta\phi^{\mu\nu}}}\Big\rangle\simeq-\frac{i}{2}\epsilon_{\sigma\nu}\langle\phi_{\mu}^{\sigma}\rangle\delta(x-x'),\label{eq: second order result}
\end{equation}
where by continuing the series this leads to 
\begin{equation}
\Big\langle W\frac{{\delta S}}{{\delta\phi^{\mu\nu}}}\Big\rangle\simeq ie^{i\langle\phi_{\mu\nu}\rangle e^{\mu}\delta x^{\nu}}\delta(x-x')=i\langle e^{i\phi_{\mu\nu}e^{\mu}\delta x^{\nu}}\rangle\delta(x-x').\label{eq: continuing the series}
\end{equation}
The input of an expansion of the field $\phi$ results in the expectation
of an operator with the form of the displacement operator. 

It is now important to understand the form the fields in the expansion
in $D(\alpha)$. The Wilson loop is a form of the Stokes' law \cite{key-29}
and 

\begin{equation}
-iln(W)=\int_{\epsilon}\partial_{\alpha}\phi_{\mu\nu}e^{\mu}d\epsilon^{\alpha\nu}.\label{eq: Fisk}
\end{equation}
In vacuum the canonical $h_{++}$ and $h_{\times\times}$ polarizations
obey $\square h_{++}=\square h_{\times\times}=0$ \cite{key-17}.
The longitudinal modes due to $R^{2}$ terms obeys \cite{key-15,key-16}

\begin{equation}
\square h_{c}=m^{2}h_{c},\label{eq: ubbidisce}
\end{equation}
where the mass is a topologically induced mass. The longitudinal $h_{c}=\phi^{2}$
then defines the equation \cite{key-15,key-16,key-20}

\begin{equation}
\square\phi_{mu\nu}=m^{2}\phi_{mu\nu}\label{eq: quadratello}
\end{equation}
where a Lorenz gauge sets terms with $\square\phi=0$ \cite{key-15,key-16,key-20}.
This term plays a role similar to the Helmholtz potential in electromagnetism

\begin{equation}
\Phi=\frac{1}{{4\pi\epsilon_{0}}}\int_{V}d^{3}r\frac{{\rho({\vec{r}})}}{{|{\vec{r}}-{\vec{r}}'|}},\label{eq: Helmholtz potential}
\end{equation}
but in the case of $f(R)$ theories it results an effective potential
through the identifications \cite{key-15,key-20}

\begin{equation}
\begin{array}{ccccc}
\Phi\rightarrow f'(R) &  & \textrm{and } &  & \frac{dV}{d\Phi}\rightarrow\frac{2f(R)-Rf'(R)}{3}\end{array}\label{eq: identifica}
\end{equation}
which give a Klein - Gordon equation for the effective $\Phi$ scalar
field \cite{key-15,key-20}

\begin{equation}
\square\Phi=\frac{dV}{d\Phi}.\label{eq: KG2}
\end{equation}
The $\phi_{{\mu\nu}}$ which physically contributes to the Wilson
integral has a source term which is the topological mass.

\section{Potential imprint in Cosmic Microwave Background}

We recall that CMB is thermal radiation filling the observable Universe
almost uniformly \cite{key-21,key-22}. Precise measurements of CMB
are fundamental for cosmology, because any viable proposed model of
the Universe must explain this radiation. The CMB has a thermal black
body spectrum at a temperature of $\sim2.7\mbox{ }K$ \cite{key-21,key-22}.
At the present time, the best available data on CMB arise from the
Planck satellite \cite{key-21,key-22}, which has produced detailed
all-sky observations over nine frequency bands between 30 and 857
GHz. According to the data, subtle fluctuations in CMB temperature
were imprinted on the deep sky during the recombination era, i.e.
when the Universe was about $370,000$ years old. That imprint reflects
ripples that arose from the early era, at about $10^{-30}$ seconds
after the initial singularity. It is a common opinion that such ripples
should give rise to the current cosmic structure of galactic clusters
and dark matter. 

The Planck satellite works within the Solar System and to take into
account weak potential effects on CMB by relic massive GWs we can
use the weak field approximation (the linearized theory). In the linearized
theory, the standard expansion $g_{\mu\nu}=\eta_{\mu\nu}+h_{\mu\nu}$
with \textquotedblleft{}small\textquotedblright{} $h_{\mu\nu}$ is
performed in an asymptotically Cartesian coordinate system. This frame
is the proper reference frame of a local observer, which we assume
to be located in A within the Solar System. In other words, we assume
that the space-time within the Solar System is locally flat with respect
to the global distribution of CMB. Our goal is to understand how relic
massive GWs perturb the trajectories of CMB photons between A and
B. The global effect results a particular \emph{gravitational lensing}
\cite{key-30} due to relic massive GWs. Some clarifications are needed
concerning this issue. In our linearized approach, gravitational lensing
can be described in the local Lorentz frame perturbed by the first
order post-Newtonian potential. Hence, one can define a refractive
index \cite{key-30,key-31}

\begin{equation}
n\equiv1+2|V|.\label{eq: indice rifrazione}
\end{equation}
In the usual \emph{Geometrical Optics}, the condition $n>1$ implies
that the light in a medium is slower than in vacuum \cite{key-32}.
Then, the effective speed of light in a gravitational field is expressed
by \cite{key-30,key-31,key-32}

\begin{equation}
v=\frac{1}{n}\approx1-2|V|.\label{eq: vel mezzo}
\end{equation}
Thus, one can obtain the Shapiro delay \cite{key-33} by integrating
over the optical path between the source and the observer:

\begin{equation}
\int_{source}^{observer}2|V|dl.\label{eq: sentiero}
\end{equation}
The situation is analogous to the prism \foreignlanguage{italian}{\cite{key-32}}.

\subsection{Gravitational lensing in the direction of the propagating gravity
wave}

For a sake of simplicity, we assume that A and B are both located
in the direction of the propagating massive GW which we assume to
be the $z$ direction.

By using the proper reference frame of a local observer the time coordinate
$x_{0}$ is the proper time of the observer A and the spatial axes
are centered in A. In the special case of zero acceleration and zero
rotation the spatial coordinates $x_{j}$ are the proper distances
along the axes and the frame of the local observer reduces to a local
Lorentz frame \foreignlanguage{italian}{\cite{key-17}}. The line
element is \foreignlanguage{italian}{\cite{key-17}}

\begin{equation}
ds^{2}=-(dx^{0})^{2}+\delta_{ij}dx^{i}dx^{j}+O(|x^{j}|^{2})dx^{\alpha}dx^{\beta}.\label{eq: metrica local lorentz}
\end{equation}
The connection between Newtonian theory and linearized gravity is
well known \cite{key-13}

\begin{equation}
g_{00}=1+2V,\label{eq: potenziale newtoniano}
\end{equation}
being $V$ the Newtonian potential. Let us consider the interval for
photons propagating along the $z$ -axis

\begin{equation}
ds^{2}=g_{00}dt^{2}+dz^{2}.\label{eq: metrica osservatore locale}
\end{equation}
The condition for a null trajectory ($ds=0$) gives the coordinate
velocity of the photons

\begin{equation}
v_{p}^{2}\equiv(\frac{dz}{dt})^{2}=1+2V(t,z),\label{eq: velocita' fotone in gauge locale}
\end{equation}
which to first order is well approximated by

\begin{equation}
v_{p}\approx[1+V(t,z)].\label{eq: velocita fotone in gauge locale 2}
\end{equation}
Knowing the coordinate velocity of the photon, the propagation time
for its traveling between A and B, which corresponds to the proper
distance AB in presence of the graviton, can be defined:

\begin{equation}
T_{1}(t)=\int_{z_{A}}^{z_{B}}\frac{dz}{v_{p}}\approx T-\int_{0}^{T}V(t',z)dz,\label{eq:  tempo di propagazione andata gauge locale}
\end{equation}
where $T$ represents the uniform propagation time of the photon between
A and B (i.e the proper distance between A and B in natural units)
as if it were moving in a flat space-time, i.e. in absence of GW,
and $t'$ is the delay time which corresponds to the unperturbed photon
trajectory: 

\begin{flushleft}
\begin{equation}
t'=t-(T-z)\label{eq: ritardo}
\end{equation}
(i.e. $t$ is the time at which the photon arrives in the position
$T$, so $T-z=t-t'$). In order to compute $T_{1}$ we need to know
the Newtonian potential $V(t,z)$ which is generated by the massive
GW. We recall that the effect of the gravitational force on test masses
is described by the equation
\par\end{flushleft}

\begin{equation}
\ddot{x^{i}}=-\widetilde{R}_{0k0}^{i}x^{k},\label{eq: deviazione geodetiche}
\end{equation}
which is the equation for geodesic deviation in this frame \foreignlanguage{italian}{\cite{key-17}}.
$\widetilde{R}_{0k0}^{i}$ is the linearized Riemann tensor \foreignlanguage{italian}{\cite{key-17}.}

\selectlanguage{italian}%
On the other hand, with an opportune choice of the Lorenz gauge, the
linearization process of $f(R)$ theories which generates the third
longitudinal mode $h_{c}=h_{c}(t-v_{G}z)$ enables a conformally flat
line element \cite{key-15,key-16,key-20}

\selectlanguage{english}%
\begin{equation}
ds^{2}=[1+h_{c}(t-v_{G}z)](-dt^{2}+dz^{2}+dx^{2}+dy^{2}).\label{eq: metrica puramente scalare}
\end{equation}
$v_{G}$ represents the group velocity of the massive GW. In fact,
the velocity of every standard massless tensorial mode $\bar{h}_{\mu\nu}$
is the light speed $c$, but the dispersion law for the modes of $h_{c}$
is that of a massive field which can be discussed like a wave-packet\foreignlanguage{italian}{\cite{key-15,key-16,key-20}}.
Also, the group-velocity of a wave-packet of $h_{c}$ centered in
$\overrightarrow{p}$ is \foreignlanguage{italian}{\cite{key-15,key-16,key-20}}

\begin{equation}
\overrightarrow{v_{G}}=\frac{\overrightarrow{p}}{\omega},\label{eq: velocita' di gruppo}
\end{equation}
which is exactly the velocity of a massive particle with mass $m$
(see Eq. (\ref{eq: ubbidisce})) and momentum $\overrightarrow{p}$.
This group-velocity is function of both of the mass and frequency
of the wave-packet \foreignlanguage{italian}{\cite{key-15,key-16,key-20}}

\begin{equation}
v_{G}=\frac{\sqrt{\omega^{2}-m^{2}}}{\omega}.\label{eq: velocita' di gruppo 2}
\end{equation}
Even if the coordinates (\ref{eq: metrica local lorentz}) are different
from the coordinates (\ref{eq: metrica puramente scalare}), we recall
that the linearized Riemann tensor is \emph{gauge invariant} \foreignlanguage{italian}{\cite{key-17}.
Hence, we can calculate it directly from Eq. }(\ref{eq: metrica puramente scalare})\foreignlanguage{italian}{.
}Following \foreignlanguage{italian}{\cite{key-16}} it is:

\begin{equation}
\widetilde{R}_{\mu\nu\alpha\beta}=\frac{1}{2}\{\partial_{\mu}\partial_{\beta}h_{\alpha\nu}+\partial_{\nu}\partial_{\alpha}h_{\mu\beta}-\partial_{\alpha}\partial_{\beta}h_{\mu\nu}-\partial_{\mu}\partial_{\nu}h_{\alpha\beta}\},\label{eq: riemann lineare}
\end{equation}
that, in the case eq. (\ref{eq: metrica puramente scalare}), begins
\foreignlanguage{italian}{\cite{key-16}}

\begin{equation}
\widetilde{R}_{0\gamma0}^{\alpha}=\frac{1}{2}\{\partial^{\alpha}\partial_{0}h_{c}\eta_{0\gamma}+\partial_{0}\partial_{\gamma}h_{c}\delta_{0}^{\alpha}-\partial^{\alpha}\partial_{\gamma}h_{c}\eta_{00}-\partial_{0}\partial_{0}h_{c}\delta_{\gamma}^{\alpha}\};\label{eq: riemann lin scalare}
\end{equation}
the different elements are (only the non zero ones will be written)
\foreignlanguage{italian}{\cite{key-16}}:

\begin{equation}
\partial^{\alpha}\partial_{0}h_{c}\eta_{0\gamma}=\left\{ \begin{array}{ccc}
\partial_{t}^{2}h_{c} & for & \alpha=\gamma=0\\
\\
-\partial_{z}\partial_{t}h_{c} & for & \alpha=3;\gamma=0
\end{array}\right\} \label{eq: calcoli}
\end{equation}

\begin{equation}
\partial_{0}\partial_{\gamma}h_{c}\delta_{0}^{\alpha}=\left\{ \begin{array}{ccc}
\partial_{t}^{2}h_{c} & for & \alpha=\gamma=0\\
\\
\partial_{t}\partial_{z}h_{c} & for & \alpha=0;\gamma=3
\end{array}\right\} \label{eq: calcoli2}
\end{equation}

\begin{equation}
-\partial^{\alpha}\partial_{\gamma}h_{c}\eta_{00}=\partial^{\alpha}\partial_{\gamma}h_{c}=\left\{ \begin{array}{ccc}
-\partial_{t}^{2}h_{c} & for & \alpha=\gamma=0\\
\\
\partial_{z}^{2}h_{c} & for & \alpha=\gamma=3\\
\\
-\partial_{t}\partial_{z}h_{c} & for & \alpha=0;\gamma=3\\
\\
\partial_{z}\partial_{t}h_{c} & for & \alpha=3;\gamma=0
\end{array}\right\} \label{eq: calcoli3}
\end{equation}

\begin{equation}
-\partial_{0}\partial_{0}h_{c}\delta_{\gamma}^{\alpha}=\begin{array}{ccc}
-\partial_{z}^{2}h_{c} & for & \alpha=\gamma\end{array}.\label{eq: calcoli4}
\end{equation}
By putting these results in Eq. (\ref{eq: riemann lin scalare}) one
gets \foreignlanguage{italian}{\cite{key-16}}

\begin{equation}
\begin{array}{c}
\widetilde{R}_{010}^{1}=-\frac{1}{2}\ddot{h}_{c}\\
\\
\widetilde{R}_{010}^{2}=-\frac{1}{2}\ddot{h}_{c}\\
\\
\widetilde{R}_{030}^{3}=\frac{1}{2}\square h_{c}.
\end{array}\label{eq: componenti riemann}
\end{equation}
Let us put Eq. (\ref{eq: ubbidisce}) in the third of Eqs. (\ref{eq: componenti riemann}).
We obtain \foreignlanguage{italian}{\cite{key-16}}

\begin{equation}
\widetilde{R}_{030}^{3}=\frac{1}{2}m^{2}h_{c},\label{eq: terza riemann}
\end{equation}
which shows that the field is not transversal. In fact, Eq. (\ref{eq: deviazione geodetiche})
implies \foreignlanguage{italian}{\cite{key-16}}

\begin{equation}
\ddot{x}=\frac{1}{2}\ddot{h}_{c}(t-v_{G}z)x,\label{eq: accelerazione mareale lungo x}
\end{equation}

\begin{equation}
\ddot{y}=\frac{1}{2}\ddot{h}_{c}(t-v_{G}z)y\label{eq: accelerazione mareale lungo y}
\end{equation}
and 

\begin{equation}
\ddot{z}=-\frac{1}{2}m^{2}h_{c}(t-v_{G}z)z.\label{eq: accelerazione mareale lungo z}
\end{equation}
Therefore the effect of the mass is exactly the generation of a \textit{longitudinal}
force (in addition to the transverse one). Note that in the limit
$m\rightarrow0$ the longitudinal force vanishes.

Equivalently we can say that there is a gravitational potential \cite{key-16,key-17}:

\begin{equation}
V(\overrightarrow{r},t)=-\frac{1}{4}\ddot{h}_{c}(t-v_{G}z)[x^{2}+y^{2}]+\frac{1}{2}m^{2}\int_{0}^{z}h_{c}(t-v_{G}a)ada,\label{eq:potenziale in gauge Lorentziana generalizzato}
\end{equation}
which generates the tidal forces, and that the motion of the test
mass is governed by the Newtonian equation \cite{key-16,key-17}

\begin{equation}
\ddot{\overrightarrow{r}}=-\bigtriangledown V.\label{eq: Newtoniana}
\end{equation}
Now, we can use Eq. (\ref{eq:potenziale in gauge Lorentziana generalizzato})
to compute $T_{1}$ in Eq. (\ref{eq:  tempo di propagazione andata gauge locale}).
We get

\begin{equation}
T_{1}(t)\approx T-\int_{0}^{T}V(t',z)dz=T-\frac{1}{2}m^{2}\int_{0}^{T}dz\int_{0}^{z}h_{c}(t'-v_{G}a)ada.\label{eq: T1 nuovo}
\end{equation}
Thus, the variation of the proper distance between A and B from its
unperturbed value $T$ which is due by the presence of the massive
GW $h_{c}$ is

\begin{equation}
\begin{array}{c}
\delta T_{1}(t)\approx\frac{1}{2}m^{2}\int_{0}^{T}dz\int_{0}^{z}h_{c}(t-T+a-v_{G}a)ada=\\
\\
=\frac{1}{4}m^{2}\int_{0}^{T}h_{c}(t-v_{G}z-T+z)dz-\frac{1}{4}m^{2}\int_{0}^{T}\int_{0}^{z}h_{c}'(t-T+a-v_{G}a)z^{2}dadz.
\end{array}\label{eq: delta T1}
\end{equation}
Introducing the Fourier transform of $h_{c}$ defined by 

\begin{equation}
\tilde{h}_{c}(\omega)=\int_{-\infty}^{\infty}dth_{c}(t)\exp(i\omega t),\label{eq: trasformata di fourier}
\end{equation}
eq. (\ref{eq: delta T1}) can be integrated in the frequency domain
by using the Fourier translation and derivation theorems

\begin{equation}
\frac{\delta\tilde{T}_{1}(\omega)}{T}=\Upsilon(\omega)\tilde{h}_{c}(\omega),\label{eq: segnale}
\end{equation}
where 
\begin{equation}
\begin{array}{c}
\Upsilon(\omega)=\frac{1}{4}m^{2}\frac{\exp(i\omega T)}{i\omega T(v_{G}-1)}\{\exp i\omega T(v_{G}-1)-1+\\
\\
+\frac{1}{i\omega(v_{G}-1)}\left[T^{2}\exp i\omega T(v_{G}-1)-2T\exp i\omega T(v_{G}-1)+2\exp i\omega T(v_{G}-1)-1\right]-\frac{T^{3}}{3}\},
\end{array}\label{eq: longitudinal response function}
\end{equation}
is the longitudinal response function for relic gravitons.

In order to use eqs. (\ref{eq: segnale}) and (\ref{eq: longitudinal response function})
we recall that relic gravitons represent a stochastic background \cite{key-18,key-20}.
Hence, one has to use average quantities \cite{key-18,key-20}. The
well known equation for the characteristic amplitude \cite{key-18},
adapted for the third component of GWs can be used \cite{key-20}:

\begin{equation}
h_{cc}(f)\simeq1.26*10^{-18}(\frac{1Hz}{f})\sqrt{h_{100}^{2}\Omega_{gw}(f)},\label{eq: legame ampiezza-spettro}
\end{equation}
obtaining, for example at 100 HZ and taking into account the bound
(\ref{eq: spettro-1}), 

\begin{equation}
h_{cc}(100Hz)\simeq1.7*10^{-26}.\label{eq: limite per lo strain}
\end{equation}
Considering a graviton propagating with a speed of $v_{G}=0.999$
(ultra-relativistic case), if we insert these values in eqs. (\ref{eq: segnale})
and (\ref{eq: longitudinal response function}) we get $\Upsilon(\omega)\approx0.02$
and $\delta\tilde{T}_{1}\approx3.4*10^{-25}m$ for a proper distance
between A and B of unperturbed value $T=1km.$ The situation is different
for a speed of $0.9$ (relativistic case). In that case one has $\Upsilon(\omega)\approx0.19$
and $\delta\tilde{T}_{1}\approx3.4*10^{-24}m.$ For a speed of $0.1c$
(non relativistic case) we have $\Upsilon(\omega)\approx0.99$ and
$\delta\tilde{T}_{1}\approx1.6*10^{-23}m.$ The situation is better
at lower frequencies. For $f=10Hz$ eq. (\ref{eq: legame ampiezza-spettro})
gives $h_{cc}\simeq1.7*10^{-25}$. The response functions result practically
unchanged, therefore we gain an order of magnitude, i.e. $\delta\tilde{T}_{1}\approx3.4*10^{-24}m$
for $v_{G}=0.999$, $\delta\tilde{T}_{1}\approx3.4*10^{-23}m$ for
$v_{G}=0.9$, and $\delta\tilde{T}_{1}\approx1.6*10^{-22}m$ for $v_{G}=0.1.$

Here we discussed the variation of the photons' paths in the $z$
direction which is the direction of the propagating relic GW. Clearly,
analogous effects, which are due by the transverse effect of the GW
(eqs. (\ref{eq: accelerazione mareale lungo x}) and (\ref{eq: accelerazione mareale lungo y})),
are present in the $x$ and $y$ directions. Thus, eqs. (\ref{eq:potenziale in gauge Lorentziana generalizzato})
and (\ref{eq: vel mezzo}) can be used to discuss the general gravitational
lensing in our model. We developed the complete computation in the
$z$ direction, the extension to the $x$ and $y$ directions is similar. 

The global effect of these variations of the photons' paths in CMB
should be analogous to the effect of water waves, which, in focusing
light, create optical caustics which are commonly seen on the bottom
of swimming pools. 

We stress that there are indications in the literature, see for instance
\cite{key-64}, that there is no amplification for $f(R)$ if compared
with general relativity, while in this paper we claim the amplification
\cite{key-44}. The key point here is the following. The ordinary
transverse strain due to the scalar field in $f(R)$ theories is,
in general, even lower with respect to the standard transverse strain
in general relativity. On the other hand, due to the presence of the
mass, in $f(R)$ theories the third scalar polarization admits also
a longitudinal strain. In this case, the correspondent longitudinal
response function, i.e. eq. (\ref{eq: longitudinal response function})
in this paper, is frequency dependent. Thus, at high frequencies,
the total signal can, in principle, be higher in f(R) theories with
respect to general relativity. This is also in agreement with the
results in \cite{key-7,key-15,key-16}.

\section{Chaos and Relativity in Orbital and Optical Systems}

The consequences of GWs form $f(R)$ theories are observable fingerprints
on the structure of the universe. Massive GWs will act as lenses which
generate caustics in the motion of light and other particle fields.
These caustics will then have measurable influences on the CMB or
upon the distribution of galaxies in the universe out to $z=1$ and
beyond. The following looks at the issue of how general relativity
can amplify chaotic dynamics, and further can amplify optical chaos.
This is illustrated in a three body problem, and in an elementary
optical model. This digression into another aspect of relativity is
meant as a way to set up analysis for the phenomenology of massive
gravity waves. This illustrates how to proceed through the examination
of elementary systems. The extension to more complex structures, such
as a many body problem of galaxies and dark matter, will require numerical
methods.

One of the early tests of general relativity was that it predicted
the perihelion precession in the orbit of Mercury \cite{key-17}.
This is a departure from Newtonian gravity that is largely post-Newtonian,
or first order or to $O(1/c^{2})$. These general relativistic corrections
are completely integrable and there is no chaotic dynamics associated
with them. In a three body problem, with a large central mass, a larger
distant mass which is treated as Newtonian and a smaller satellite
with $O(1/c^{2})$ relativistic departures, will exhibit chaotic dynamics
in the small body. The additional relativistic corrections will interplay
with the irregular chaotic dynamics and are shown below to contribute
to a Lyapunov exponent \cite{key-34}. In effect a Lyapunov exponent
$\lambda=log(\Lambda)$ will have a relativistic correction $\Lambda=\Lambda_{0}+\Lambda(O(c^{-2}))$,
and this correction then amplifies the chaotic behavior of the system.
This is extended to optical systems. Einstein lenses \cite{key-35}
are a Newtonian gravitational phenomenon, and general relativistic
corrections to $O(1/c^{2})$ are minor, for the impact parameter on
such a gravitating body is too small to be observationally significant.
Yet for a complex Einstein lens, say analogous to a compound lens
due to smaller scale clumping of matter, a light ray may have a succession
of small angular deviations. These angles of deviation will have a
compounding effect similar to angle deviations of a particle in an
arena. This will result in increasingly complex optical caustics which
in analogue with chaos are difficult to predict. This is further compounded
if the gravitating clumps of matter are difficult to observe directly,
such as with dark matter \cite{key-36}. In a manner similar to the
case with orbital dynamics general relativistic corrections may also
enhance this optical chaos or turbulence. This Section connects two
different aspects of chaos and relativity to present issues with the
analysis of three body systems with parameterized post-Newtonian parameters.
Subtle enhancements of chaotic dynamics or the increase in a Lyapunov
exponent might be documented in such a system. This should then be
an observable characteristic of complex relativistic systems. The
optical analogue illustrates how fine detailed structure in a distribution
of matter which is an Einstein lens could influence the complexity
of optical caustics. Localized regions of large gravity fields could
then further amplify this complexity as well. This might lead to methods
for mapping any local density variation in dark matter.

We stress that the numerical values of the Lyapounov exponent $\lambda$
in general relativity are not gauge invariant, that is, they depend
on the chosen coordinate system \cite{key-62}. Therefore, for the
same dynamical system, chaotic behavior may appear in some frames,
but not in others \cite{key-62}. Following \cite{key-63}, we find
three different problems when one uses the Lyapounov exponent in general
relativity:
\begin{enumerate}
\item The reference systems have no unified time. 
\item The separation of space and time in the 4-dimensional spacetime varies
for different observers. 
\item Time and space coordinates works only for events and sometimes have
no physical meaning. 
\end{enumerate}
Consequently, we could get different values of the Lyapounov exponent
in different coordinate systems. The problem can be solved if one
uses proper time and proper distances instead \cite{key-63}. In that
case is indeed possible consider a particle, called \textquotedblleft{}observer\textquotedblright{},
moving along an orbit in the spacetime \cite{key-63}. That particle
can understand if its motion is or is not chaotic observing if the
proper distances from neighbors particles are increasing exponentially
or not with its proper time \cite{key-63}. Hence, the point in \cite{key-62}
that the Lyapounov exponent is not gauge invariant in general relativity
is correct. However, the point of this is to examine the possible
role of general relativity in the amplification of chaotic dynamics.
In effect, general relativity applies to a body close to the star
or large mass, where these gravitationally interact by Newtonian gravity
to a third body. The purpose is to illustrate how chaos in Newtonian
mechanics may be amplified if the system interacts with a semi-relativistic
system in a stronger gravity field. Within this approximation the
question concerning invariance of the Lyapunov exponent in general
relativity for the Newtonian dynamical body is a small effect. The
Lyapunov exponent applies strictly to the Newtonian part of the problem.

\subsection{General relativity to $O(1/c^{2})$ }

In general relativity the equation of motion for a test mass particle
around a fixed central mass is \cite{key-17}

\begin{equation}
\frac{{d^{2}u}}{{d\theta^{2}}}+u=\frac{{GM}}{{l^{2}}}+\frac{{3GMu^{2}}}{{c^{2}}}.\label{eq: 4.1}
\end{equation}
Here $l$ is the constant specific angular momentum. We recognize
this differential as the harmonic oscillator equation of Newtonian
mechanics with a constant force $GM/l^{2}$, plus the term $\sim(u/c)^{2}$.
The anomaly angle $\theta$ obeys the dynamical equation \cite{key-17}

\begin{equation}
\frac{{d\theta}}{{ds}}=\frac{l}{{r^{2}}}=lu^{2}\label{eq: 4.2}
\end{equation}
and 
\begin{equation}
\frac{{dt}}{{ds}}=\frac{E}{{1-2GMu/c^{2}}}\label{eq: 4.3}
\end{equation}
for $E$ the potential energy per unit mass of the particle \textquotedbl{}at
infinity\textquotedbl{} $=constant$. For $GM/c^{2}\ll1$ we may solve
this problem by perturbation methods. The solution of interest is
$O(1)$ plus $O(c^{-2})$, which would be Newton plus first order
GR correction. The expansion is carried out with the variables $u,\theta$
according to

\begin{equation}
u=u_{0}+\epsilon u_{1}+O(\epsilon^{2}),\label{eq: 4.4}
\end{equation}

\begin{equation}
\theta=\theta_{0}+\epsilon\theta_{1}+O(\epsilon^{2}).\label{eq: 4.5}
\end{equation}
Here the term $\epsilon=1/c^{2}$ gives the order of the expansion.
The differential with respect to $\theta$ to first order in $\epsilon$
is taken as 
\begin{equation}
\frac{d}{{d\theta}}\simeq\frac{d}{{d\theta_{0}}}+\epsilon\frac{d}{{d\theta_{1}}}.\label{eq: 4.6}
\end{equation}
If we input the expansion for $u$ in equation (\ref{eq: 4.4}) into
the differential equation of motion (\ref{eq: 4.1}) the following
two equations are obtained:

\begin{equation}
O(1):\frac{{d^{2}u_{0}}}{{d\theta_{0}^{2}}}+u_{0}=\frac{\kappa}{{l^{2}}},\label{eq: 4.7}
\end{equation}

\begin{equation}
O(\epsilon):\frac{{d^{2}u_{1}}}{{d\theta_{0}^{2}}}+u_{1}-3\kappa u_{0}^{2}=0.\label{eq: 4.8}
\end{equation}
The term ${{d^{2}u_{0}}/{d\theta_{0}d\theta_{1}}}=0$ since $u_{0}$
is not a function of $\theta_{1}$. Further, the term $\kappa=GM$.
The $O(1)$ differential equation (\ref{eq: 4.7}) has the solution

\begin{equation}
u_{0}=\frac{\kappa}{{l^{2}}}(1+\epsilon^{\prime}cos(\theta_{0}+\alpha)),\label{eq: 4.9}
\end{equation}
which is the standard Newtonian solution for the radial velocity for
a particle with orbital eccentricity $\epsilon^{\prime}$ and anomaly
angle $\alpha$ \cite{key-17}. Now, let us consider on the expansion
of $\theta$. We set $E=1$ and insert this into the equation for
the angular velocity equation 

\begin{equation}
\frac{{d\theta}}{{dt}}=(1-2\kappa\epsilon u)lu^{2},\label{eq: 4.10}
\end{equation}
where $1-2\kappa\epsilon u$ is the Schwarzschild transformation between
proper and standard time coordinates \cite{key-17}. This differential
equation has the two contributing parts: 
\begin{equation}
O(1):\frac{{d\theta_{0}}}{{dt}}=lu_{0}^{2}\label{eq: 4.11}
\end{equation}
 
\begin{equation}
O(\epsilon):\frac{{d\theta_{1}}}{{dt}}=2lu_{0}-2\kappa lu_{0}^{3}.\label{eq: 4.12}
\end{equation}
We are primarily concerned at this point in the solution to order
$O(\epsilon)$ for the orbit of a test mass in a GR orbit, 
\begin{equation}
\frac{{d^{2}u_{1}}}{{d\theta_{0}^{2}}}+u_{1}-3\kappa u_{0}^{2}=0\label{eq: 4.13}
\end{equation}
 where the Newtonian solution $u_{0}$ is given by equation (\ref{eq: 4.9}).
The square of $u_{0}$ in the non-homogenous term is 
\begin{equation}
u_{0}^{2}=\Big(\frac{\kappa}{{l^{2}}}\Big)^{2}\Big(1+2\epsilon^{\prime}cos(\theta_{0}+\alpha)+{\epsilon^{\prime}}^{2}cos^{2}(\theta_{0}+\alpha)\Big),\label{eq: 4.14}
\end{equation}
 which by elementary trigonometric identities is 
\begin{equation}
u_{0}^{2}=\Big(\frac{\kappa}{{l^{2}}}\Big)^{2}\Big((1-\epsilon^{\prime})+2\epsilon^{\prime}cos^{2}((\theta_{0}+\alpha)/2)+{\epsilon^{\prime}}^{2}cos^{2}(\theta+\alpha)\Big).\label{eq: 4.15}
\end{equation}
 The reason for doing this is that the solution is elementary at this
point. The first non-homogeneous term is going to give a solution
\begin{equation}
\sim\frac{\kappa^{3}}{{l^{4}}}(1-\epsilon^{\prime})(1+\epsilon^{\prime}cos(\theta_{0}+\alpha)),\label{eq: 4.16}
\end{equation}
 and the quadratic trigonometric functions determine the solution:
\begin{equation}
u_{1}=\frac{\kappa^{3}}{{l^{4}}}\Big(\big(1+\epsilon^{\prime}cos(\theta_{0}+\alpha)\big)+\frac{{2\epsilon^{\prime}}}{3}\big(cos(\theta_{0}+\alpha)-3\big)+\frac{{{\epsilon^{\prime}}^{2}}}{2}\big(cos(2(\theta_{0}+\alpha))-3\big)\Big).\label{eq: 4.17}
\end{equation}
The hard part is the perturbation of the third planet. The Jovian
planet obeys a similar dynamical equation, but where $c\rightarrow\infty$
and Newtonian dynamics is recovered as 
\begin{equation}
\frac{{d^{2}v}}{{d\theta'^{2}}}+v=\frac{{\kappa}}{{L^{2}}}.\label{eq: 4.18}
\end{equation}
Here $v=1/r_{2}$ for this additional planet, and we define $u=u_{0}+\epsilon u_{1}=1/r_{1}$.
Similarly, the angular momentum is defined by \cite{key-17} 
\begin{equation}
\frac{{d\theta'}}{{dt}}=\frac{L}{{r_{2}^{2}}}=Lv^{2}.\label{eq: 4.19}
\end{equation}
The angle $\theta'$ may exist in a different plane than $\theta$,
yet as an approximation we put both angles in the same plane of motion.
Now we need the coupling between the two bodies. We assume they are
Newtonian as 
\begin{equation}
{\vec{F}}=GMm\frac{{{\vec{r}}_{1}-{\vec{r}}_{2}}}{{|{\vec{r}}_{1}-{\vec{r}}_{2}|^{3}}},\label{eq: 4.20}
\end{equation}
which is approximately 
\begin{equation}
{\vec{F}}=\frac{{GMm}}{{r_{2}^{3}}}(1+\frac{3}{2}\frac{{{\vec{r}}_{1}\cdot{\vec{r}}_{2})}}{{r_{2}^{2}}}({\vec{r}}_{1}-{\vec{r}}_{2}).\label{eq: 4.21}
\end{equation}
 To find the distance $|{\bf r}_{1}-{\bf r}_{2}|$ we consider the
plane of the two orbits as complex valued and that the positions of
the test mass and the larger mass are give by ${\bf r}_{1}=r_{1}e^{i\theta_{1}}$
and ${\bf r}_{2}=r_{2}e^{i\theta_{2}}$ and so the distance between
the two masses is given by 
\begin{equation}
|{\bf r}_{1}-{\bf r}_{2}|^{2}=r_{1}^{2}+r_{2}^{2}-2r_{2}r_{2}cos(\theta_{1}+\theta_{2}).\label{eq: 4.22}
\end{equation}
The potential energy 
\begin{equation}
U(r_{1},r_{2})=-\frac{{GMm}}{{|{\bf r}_{1}-{\bf r}_{2}|}}\label{eq: 4.23}
\end{equation}
defines the force in equation (\ref{eq: 4.20}) by ${\bf F}=-\nabla U$.
For $r_{2}\gg r_{1}$ the denominator in the potential may then be
cast in the $u,v$ variables 
\begin{equation}
U(u_{1},u_{2})\simeq GMmv\Big(1-\Big(\frac{v}{u}\Big)^{2}+2\frac{v}{u}cos\big((\omega_{1}+\omega_{2})t\big)\Big).\label{eq: 4.24}
\end{equation}
Here $\omega_{i}=d\theta_{i}/dt$, for $i=1,2$ for the two bodies.
This is the perturbing potential for the two orbits of the bodies
in the same plane. 

The total Hamiltonian is then 
\begin{equation}
\begin{array}{c}
H=+\epsilon H_{1}^{ho}+H_{v}^{ho}+\kappa^{\prime}v\Big\{(1-\Big(\frac{v}{{u_{0}}}\Big)^{2}+\Big(\frac{v}{{u_{0}}}\Big)cos(\theta+\theta^{\prime})\Big\}\\
\\
-3\epsilon u_{0}^{2}u_{1}-\epsilon\kappa^{\prime}\Big(1-\frac{v}{u_{0}}\Big)\Big(\frac{v}{{u_{0}}}\Big)^{2}u_{1},
\end{array}\label{eq: 4.25}
\end{equation}
where the first three terms are harmonic oscillator Hamiltonians 
\begin{equation}
H_{0}^{ho}=\frac{1}{2}p_{0}^{2}+\frac{1}{2}\frac{\kappa}{{l^{2}}}u_{0}^{2},\mbox{ }H_{1}^{ho}=\frac{1}{2}p_{1}^{2}+\frac{1}{2}\frac{\kappa}{{l^{2}}}u_{1}^{2},\mbox{ }H_{v}^{ho}=\frac{1}{2}p_{v}^{2}+\frac{1}{2}\frac{\kappa^{\prime}}{{L^{2}}}v^{2}.\label{eq: 4.26}
\end{equation}
We now have two order parameters $\epsilon=1/c^{2}$, and another
$\kappa^{\prime}=Gm$, where the mass $m$ is the mass of the \textquotedbl{}Jovian\textquotedbl{}
planet. The Hamiltonian term that scales according to $\epsilon\kappa^{\prime}$
for $\theta=\theta_{0}+\epsilon\theta_{1}$ is 
\begin{equation}
H_{\epsilon\delta}\simeq-\kappa{}^{\prime}\Big(1-\frac{v}{u_{0}}\Big)\Big(\frac{v}{{u_{0}}}\Big)^{2}u_{1}\Big(1-\frac{v}{u_{0}}\Big)\Big(\frac{v}{{u_{0}}}\Big)^{2}u_{1}.\label{eq: 4.27}
\end{equation}
To compute the Lypunov exponent explicitly the gradients of the Hamiltonian
with $p_{0},\mbox{ }p_{1},\mbox{ }p_{v}$ and $u_{0},\mbox{ }u_{1},\mbox{ }v$
are first found. With $v/u_{0}\ll1$, $u_{1}\ll u_{0}$ these are
then to order $(v/u_{0})^{2}$ 
\begin{equation}
\nabla_{p_{0}}H={\dot{u}_{0}},\mbox{ }\nabla_{p_{1}}H=\epsilon{\dot{u}}_{1},\mbox{ }\nabla_{p_{v}}H={\dot{u}}_{v}\label{eq: 4.28}
\end{equation}
 
\begin{equation}
\nabla_{u_{0}}H=\frac{\kappa}{{l^{2}}}u_{0}-6\epsilon u_{0}u_{1}-\kappa^{\prime}\frac{{v^{2}}}{{u_{0}^{2}}}cos(\theta+\theta^{\prime})\label{eq: 4.29}
\end{equation}
 
\begin{equation}
\nabla_{u_{1}}H=\frac{{\epsilon\kappa}}{{l^{2}}}u_{1}-3\epsilon u_{0}^{2}-\epsilon\kappa^{\prime}\frac{{v^{2}}}{{u_{0}^{2}}}\label{eq: 4.30}
\end{equation}
 
\begin{equation}
\nabla_{v}H=\frac{\kappa^{\prime}}{{L^{2}}}v+\kappa^{\prime}\Big(1-2\frac{{v^{2}}}{{u_{0}^{2}}}+2\frac{{v}}{{u_{0}}}cos(\theta+\theta^{\prime})\Big).\label{eq: 4.31}
\end{equation}
These are the forces $F=-\nabla H$ due to the three configuration
variables $u_{0}$, $u_{1}$ and $v$. The last right hand side terms
in $\nabla_{u_{1}}H$ are dependent upon both the general relativistic
correction $O(1/c^{2})$ and the gravitational coupling with the Jovian
planet $\kappa^{\prime}$. 

We consider the change in the phase space flow 
\begin{equation}
Z+\Delta Z=(u+\Delta u,\mbox{ }p+\Delta p)\label{eq: 4.32}
\end{equation}
The change in momenta due to the perturbation from the Jovian planet
is 
\begin{equation}
\Delta p\simeq\Delta t\Big[\kappa^{\prime}\Big(1-2\frac{{v^{2}}}{{u_{0}^{2}}}+2\frac{{v}}{{u_{0}}}cos(\theta+\theta^{\prime})\Big)-\kappa^{\prime}\frac{{v^{2}}}{{u_{0}^{2}}}cos(\theta+\theta^{\prime})-\epsilon\kappa^{\prime}\frac{{v^{2}}}{{u_{0}^{2}}}\Big],\label{eq: 4.33}
\end{equation}
where the last term is a coupling of general relativistic $O(1/c^{2})$
effects and planetary perturbation. To $O(\kappa^{\prime}/c^{2})$
$\Delta u\propto\Delta p$. Define $\Delta p(t)$ to be the deviation
in momentum due to planetary perturbation, and let $\delta p(t)$
be the deviation due to the $O(1/c^{2})$ coupling term. The Lyapunov
exponent is then 
\begin{equation}
\lambda\simeq\lim_{t\rightarrow\infty}\frac{1}{t}ln\Big(\frac{{\Delta p(t)+\delta p(t)}}{{\Delta p(t_{0})}}\Big)\simeq\lim_{t\rightarrow\infty}\frac{1}{t}\Big[ln\Big(\frac{{\Delta p(t)}}{{\Delta p(t_{0})}}\Big)+\frac{{\Delta p(t_{0})\delta p(t)}}{{\Delta p(t)}}\Big]\label{eq: 4.34}
\end{equation}
so that 
\begin{equation}
\lambda\simeq\lambda_{0}+\epsilon\Big(\frac{v(t_{0})}{{u_{0}(t_{0})}}\Big)^{2}\Big],\label{eq: 4.35}
\end{equation}
with $\lambda_{0}$ defined for $\epsilon=0$. The exponential divergence
in phase space between nearby trajectories has then contribution in
addition to $\lambda_{0}$ with $Z(t)\sim Z(t_{0})e^{\lambda_{0}t}e^{\epsilon(v/u_{0})^{2}t}$.
Thus general relativity will bring about the onset of chaotic behavior,
or the breakdown of numerical unpredictability, earlier. 

This should then manifest itself in semi-relativistic systems with
three bodies. A system, such as two neutron stars in a mutual relativistic
orbit with a third companion further away and executing Newtonian
dynamics, of this sort will then have more chaotic behavior which
is amplified by general relativistic effects. This simplified model
suggests that a general parameterized post-Newtonian-multibody perturbative
theory is needed. Such a model will then be more suited for the examination
of complex general relativistic systems that include several bodies.

\subsection{$O(1/c^{2})$ Optical Corrections in Einstein Lensing}

The Einstein lensing of light is now a common observational feature
of deep space astronomy since the launch and repair of the Hubble
Space Telescope, see \cite{key-37} and references within. Here a
complex optical gravitational lensing system is discussed with some
analogues to the mechanics above. A large elliptical galaxy will have
an overall gravitational lensing effect, but there may be sub-lensing
as well if the density of dark matter exists has some variation. This
results in a type of optical turbulence, analogous to chaos. Further,
this may also be amplified by general relativistic effects. Unknown
configurations might exist with dark matter density increasing in
the vicinity of a large black hole. 

The general theory of gravitational lensing \cite{key-30,key-31}
shows that a light ray which approaches within a radius $r\gg2GM/c^{2}$
will be deflected approximately by an angle $\theta=GM/rc^{2}$. In
a more general setting the deflection of light is given by the Einstein
angular radius 
\begin{equation}
\theta_{E}=\sqrt{\frac{{4GM}}{{c^{2}}}\frac{{d_{ls}}}{{d_{l}d_{s}}}},\label{eq: 4.36}
\end{equation}
where $d_{ls},\mbox{ }d_{l},\mbox{ }d_{s}$ are the angular diameters
to the gravitational lens, the source and the distance between the
gravitational lens and the source. For $d_{ls},\mbox{ }d_{l},\mbox{ }d_{s}$
the angular diameters to the gravitational lens, the source and the
distance between the gravitational lens and the source. The condition
$d_{s}=d_{l}+d_{sl}$ obtains locally where cosmological frame dragging
is small. This theory is the weak gravitational lensing approximation,
where the deflection of light is essentially a Newtonian result \cite{key-30}.
The distance relationships are determined by $\theta d_{s}=\beta d_{s}+\alpha'd_{ls}$.
The reduced angle of deflection $\alpha(\theta)=(d_{ls}/d_{s})\alpha'(\theta)$
gives a relationship between the angles of importance $\alpha(\theta)+\beta=\theta$.

Complex distributions of matter can act similar to a compound lens
in a weak gravitational limit. However, light rays which pass close
to clumps of matter to exhibit $O(1/c^{2})$ deviations will exhibit
deviations from this linear summation. A light ray which passes through
a set of \emph{random lenses} will display caustics which are similar
to the caustics seen on the bottom of a swimming pool. Fine grained
structure in an Einstein lens can exhibit caustics which occur due
to nonlinear perturbation in the density profile of matter. This nonlinearity
in the symmetry of the lens will produce caustics which are analogous
to chaos. The occurrence of a caustic has its connections with catastrophe
theory \cite{key-38} and the onset of a fold, which is also a mechanism
for the bifurcation of vector fields in Hamiltonian chaos. 

For the position of a source $\vec{x}$, the propagation of light
along the $z$ axis from this source then reduces the visual appearance
of the object to $\vec{\xi}=(\xi_{x},\mbox{ }\xi_{y})$ along the
axis of optical propagation. The weak gravitational lensing of light
\cite{key-30} then indicates that the deflection of the appearance
of this object along the axis of optical propagation is given by 
\begin{equation}
\Delta{\vec{\xi}}=\nabla\Phi(\xi),\label{eq: 4.37}
\end{equation}
for $\xi$ the position of the image with the mass present and $\Phi(\xi)$
the gravitational potential. The difference in the vector position
of the image ${\vec{\xi}}_{i}-{\vec{\xi}}_{s}$ is the difference
between the position with the mass present and without it being present.
The potential term obeys the Poisson equation \cite{key-13} so that
\begin{equation}
\nabla^{2}\Phi=2\frac{\Sigma(\vec{\xi})}{\Sigma_{c}}\label{eq: 4.38}
\end{equation}
The integration over the direction of propagation then gives the mass
density in the plane of the image, often called the surface mass density
$\Sigma(\vec{\xi})$. The angle of deflection $\alpha$ is then determined
by the Poisson equation and the potential as 
\begin{equation}
{\vec{\alpha}}'(\vec{\xi})=\frac{{4G}}{{c^{2}}}\int\frac{{(\vec{\xi}-\vec{\xi}')\Sigma(\vec{\xi}')}}{{|\vec{\xi}-\vec{\xi}'|^{2}}}d^{2}\xi',\label{eq: 4.39}
\end{equation}
for $\Sigma(\vec{\xi})$ a mass/area density distribution in the image.
The function $\Sigma(\vec{\xi})$ plays the role of an index of refraction
based upon a mass distribution, which for a thin lens will give the
angle of deviation. For a gravitational thin lens, a weak field that
is very small compared to the optical path length, and $\Sigma(\vec{\xi})$
is a constant. The deflection angle is simply 
\begin{equation}
\alpha(\xi)=\frac{{4\pi G}}{{c^{2}}}\frac{{\Sigma(\xi)d_{ls}\xi}}{{d_{s}}}\label{eq: 4.40}
\end{equation}
where for small angles $|\vec{\xi}|=\xi=d_{l}\theta$ and 
\begin{equation}
\alpha(\xi)=\frac{{4\pi G\Sigma}}{{c^{2}}}\frac{{d_{ls}d_{l}}}{{d_{s}}}=\frac{\Sigma}{\Sigma_{c}}\theta\label{eq: 4.41}
\end{equation}
for the critical mass density $\Sigma_{c}=(c^{2}/4\pi G)(d_{s}/d_{ls}d_{l})$.
This is the minimal mass density which might be distributed in the
area of an Einstein ring \cite{key-39}. For a more complex arrangements
of gravitational lenses, such as large density nonlinearities, the
mass density $\Sigma$ has a general form 
\begin{equation}
\Sigma(\xi)=\int dz\sigma(d_{l}\xi_{x},\mbox{ }d_{l}\xi_{l},\mbox{ }z)\label{eq: 4.42}
\end{equation}
The position of the image then plays the role of the vector $\vec{r}$
in equations (\ref{eq: 4.1})-(\ref{eq: 4.2}) and beyond in the above
discussion. Let the reciprocal of the vector norm $|\vec{\xi}|=1/\nu$
plays the role of $u$. The analogue of the Newtonian equation of
motion in equation 1 with $c~\rightarrow\infty$ is then 
\begin{equation}
\Delta\nu=\frac{\kappa}{{(c\xi)^{2}}},\label{eq: 4.48}
\end{equation}
for $\xi$ the impact parameter. Now the Newtonian description of
gravitational lens deflection has the effective photon angular momentum
per mass term $j=(c\xi)^{-1}$. The general relativistic extension
of this equation is then 
\begin{equation}
\Delta\nu=\frac{\kappa}{{(c\xi)^{2}}}+\frac{{3\kappa\nu^{2}}}{{c^{2}}}.\label{eq: 4.49}
\end{equation}
To order expansions the analogue of equations (\ref{eq: 4.7}) and
(\ref{eq: 4.8}) are 
\begin{equation}
\Delta\nu_{0}=\frac{\kappa}{{(c\xi)^{2}}},\label{eq: 5.50}
\end{equation}
 
\begin{equation}
\Delta\nu_{1}=\frac{{3\kappa\Delta\nu_{0}^{2}}}{{c^{2}}},\label{eq: 5.51}
\end{equation}
for $u_{0}$ the reciprocal of $\xi$. The term $\kappa=GM$ for a
general distribution in the plane of the Einstein ring is $\kappa/\xi\simeq4\pi G\Sigma d_{ls}d_{l}/d_{s}$,
which reproduces the Einstein ring case in the first approximation.
The second order term is $\Delta\nu=\Delta\xi/\xi^{2}$ and $\alpha\simeq2\Delta\xi/\xi$,
which reproduces the weak field gravity lens result. The $O(1/c^{2})$
correction gives an effective general relativistic correction term
\begin{equation}
\Delta\alpha\simeq\frac{{3\kappa d_{l}^{2}\theta^{2}}}{{c^{2}}}\Big(\frac{{\Sigma}}{{\Sigma_{c}}}\Big)^{2}\label{eq: 5.52}
\end{equation}
This correction term is not likely to be detected directly by extra-galactic
sources, such as the dark matter lensing of light by the Abell galaxy
cluster \cite{key-40}.

\subsection{Optical Chaos }

Just as the $O(1/c^{2})$ correction to Newtonian dynamics enhanced
chaotic dynamics, or contributed to a Lyapunov exponent, we might
expect a similar amplification of optical chaos, or the statistical
appearance of caustics by the clumping of matter. Small local region
where gravitating mass is clumped together will result in the deviation
of the light ray by some small angle $\delta\theta$, which is an
\emph{error} in computing the subsequent tracing of the ray. With
a succession of $n$ such small deviations the first angle deviation
is amplified by $\simeq2^{n}\delta\theta_{1}$, the next by $\simeq2^{n+1}\delta\theta_{2}$,
where for large $n$ and $\theta_{i}\simeq\theta$ $\forall i$ the
total angular error in computing a ray trace will be approximately
$2^{n+1}\delta\theta$. This is analogous to the arena problem of
computing the trajectory of a ball. 

The vector $\vec{\xi}$ describes the visual appearance of a distant
object along the axis of propagation. This vector describes the deformation
of a wave front by the lensing action of the intervening gravitating
body. The gravitational lens is usually considered as a symmetric
lens \cite{key-30}, but nature may provide local clumping of material
which introduce some chaos in the ray tracing. Further, the over all
gravitating lens may be sufficient enough to produce small $O(1/c^{2})$
relativistic deviations from a purely Newtonian lensing. Above the
formula for this relativistic deviation is given. What is then needed
is an analogue to Lyapunov exponent for the classical unpredictability
of a ray trace due to Newtonian gravitational sources. A multiple
set of ray tracings is then a description of the deformation of an
electromagnetic wave front, and perturbations on the vector $\vec{\xi}$.
In what follows such a development is presented to describe the chaotic
perturbation of this vector. 

The propagation of a plane electromagnetic wave front is given by
$\psi({\vec{r}})=\psi_{0}e^{i{\vec{k}}\cdot{\vec{r}}-\omega t}$ .
The occurrence of a gravitational lens perturbs the the wave front
according to 
\begin{equation}
\psi^{\prime}({\vec{r}})=\big(\chi({\vec{r}})e^{i\phi({\vec{r}})}\big)\psi({\vec{r}}).\label{eq: 5.53}
\end{equation}
Here the $\phi({\vec{r}})$ is the change in the wave front phase
and $\chi({\vec{r}})$ is the change in the wave front amplitude.
The vectors describing the visual appearance of the image are ${\vec{\xi}}=\nabla_{r_{||}}\phi({\vec{r}})$
for $r_{||}$ coordinate directions along the wave front. This means
that $\Delta\phi(\xi)=\Phi(\xi)$, which is a Poisson equation \cite{key-13}.
The phase deviations are caused by an effective index of refraction
\cite{key-30,key-31} in the Newtonian limit, and the gravitational
potential is the source in the Poisson equation. A Gaussian random
distribution of sources results in the second order structure function
\begin{equation}
D_{\phi}({\vec{\rho}})=\langle|\phi({\vec{r}})-\phi({\vec{r}}-{\vec{\rho}})|^{2}\rangle.\label{eq: 5.54}
\end{equation}
The vector ${\vec{\rho}}={\vec{\xi}}+{\vec{z}}$, where ${\vec{\xi}}={\vec{r}}-{\vec{\rho}}$,
so $D_{\phi}({\vec{\rho}})$ is a phase variance between two different
direction in the aperture plane. 

The phase terms obey a Poisson equation, where some distribution of
sources is present. For optical perturbations compatible with the
second order structure function the gravitating perturbations are
in a Gaussian distribution, where Gaussian distributions of perturbing
sources means that the equation (\ref{eq: 4.38}) becomes 
\begin{equation}
\nabla^{2}\Phi=2\frac{\Sigma(\vec{\xi})}{\Sigma_{c}}+\frac{1}{{4\pi}}\frac{\mu}{{(\sqrt{2\pi}\sigma)^{3}}}\prod_{i}e^{-\xi_{i}^{2}/2\sigma^{2}},\label{eq: 5.55}
\end{equation}
where each $\xi_{i}\ll\xi$ . Each one of these sources gives a solution

\begin{equation}
\phi=(\mu/4\pi)(1/r)Erf(r/\sqrt{2\sigma}),\label{eq: centrata}
\end{equation}
for the variable $\xi_{i}=r$, and the solution converges to a point
course in the limit $\sigma\rightarrow0$. Each of these perturbing
changes on the aperture vector is due to a succession of matter clumps.
A photon which passes close to each clump is modeled as having its
angle deviated, and its path is then stochastically deviated away
from a path given by equation (\ref{eq: 5.55}). The small angle of
deviation for ${\vec{\alpha}}(\vec{\xi})\rightarrow{\vec{\alpha}}(\vec{\xi})+\delta{\vec{\alpha}}(\vec{\xi})$
is determined by the Gaussian distribution as 
\begin{equation}
\delta{\vec{\alpha}}(\vec{\xi})=\frac{{4G}}{{c^{2}}}\int\prod_{i}\frac{{(\vec{\xi}_{i}-\vec{\xi}'_{i})\rho(\vec{\xi}'_{i})}}{{|\vec{\xi}-\vec{\xi}_{i}'|^{2}}}d^{2}\xi_{i},\label{eq: 5.56}
\end{equation}
for $\rho(\vec{\xi}_{i})=(1/4\pi)(\mu/(\sqrt{2\pi}\sigma)^{3})e^{-\xi_{i}^{2}/(2\sigma^{2})}.$
For simplicity the angle of deviation $\delta{\vec{\alpha}}(\vec{\xi})$
will now be treated as a scalar and with $\xi\gg\xi_{i}$ the angle
deviation is 
\begin{equation}
\delta\alpha\simeq\kappa\int\prod_{i}\xi^{-1}\rho(\xi_{i}')d\xi_{i}'\label{eq: 5.57}
\end{equation}
such that $\alpha\simeq\langle\xi^{-1}\rangle$. This is a partition
function analogous to that in the Ising model \cite{key-41}, but
here instead of a set of spins that exist in space there are stochastic
angle changes in a ray trace of light. In this particular model these
stochastic angle changes are assumed to be on average the same. 

This partition function can be demonstrated to be similar to the Ising
model. For the variation in the stochastic variable $\delta\xi_{j}=\xi_{j}-\xi_{j-1}$
in the exponent, the product of any two variations vanish $\delta\xi_{j}\delta\xi_{j}\simeq0$,
so that 
\begin{equation}
\xi_{i-1}\xi_{j-1}+\xi_{i}\xi_{j}=2\xi_{i-1}\xi_{j}.\label{eq: 5.58}
\end{equation}
for $i=j$ the sum of these stochastic variables is 
\begin{equation}
\frac{{1}}{{2}}\sum_{j=0}^{n-1}\xi_{j}\xi_{j}=\sum_{j=1}^{(n-1)/2}\xi_{j-1}\xi_{j}-\frac{1}{2}(\xi_{0}^{2}+\xi_{n}^{2}).\label{eq: 5.59}
\end{equation}
This means there exist additional \emph{endpoint terms} which do not
conform to the Ising type of construction. However, for a large enough
$n$ this error should be minimal. The expectation is approximately
\begin{equation}
\langle\xi\rangle\simeq\frac{1}{\sqrt{2\pi\sigma}}\int\prod_{j=1}^{(n-1)/2}d\xi_{j}\xi^{-1}exp\big(-\xi_{j-1}\xi_{j}\beta\big),\label{eq: 5.60}
\end{equation}
for $\beta=1/\sigma^{2}$. $\beta$ is analogous to the Boltzmann
factor, which is determined by the scale at which matter is lumped
together. A correlation length scale is 
\begin{equation}
\lambda^{2}\simeq1/log(tanh\beta),\label{eq: 5.61}
\end{equation}
which for $\beta\ll1$, or equivalently for large $\sigma$ is $\lambda\simeq\sigma$.
This approximate formula is a ray trace path analogue of the Lyapunov
exponent in time, which determines a length $\lambda$ where the prediction
of a ray trace breaks down. This also illustrates that this breakdown
of ray tracing occurs on a scale comparable to the length scale of
the perturbing. This loss of ray trace prediction is manifested deformations
of the angle deviation across the aperture distance, or deviations
in the symmetry of an Einstein ring. 

The goal now is to determine if there are enhancements of optical
chaos, analogous to optical turbulence in the Earth's atmosphere,
due to $O(1/c^{2})$ corrections. To examine this we consider the
angle of deviation due to Newtonian gravity, optical path length turbulence,
and relativistic corrections as conjugate to action variables $J,\mbox{ }J',\mbox{ }J"$,
and the path is given in a classical setting by the action 
\begin{equation}
(H+H^{\prime}+H^{\prime\prime})dt=(Jd\alpha(\xi)+J^{\prime}d\alpha^{\prime}(\delta\xi)+J^{\prime\prime}d\alpha^{\prime\prime}(\xi_{1}).\label{eq: 5.62}
\end{equation}
For the angular momentum variables $J\simeq J^{\prime}\simeq J^{\prime\prime}$
in this equation then the action is entirely governed by the angle
deviation, which for $d\alpha=(d\alpha/dt)dt$ expresses this as a
principle of least time. Just as in the case of planetary motion.
The angle of deviation due to clumpiness of matter is approximated
as 
\begin{equation}
\delta\alpha_{c}\simeq\kappa exp(\xi^{2}log(n)\beta),\label{eq: 5.63}
\end{equation}
for $n$ regions of matter or dark matter clumping. The region where
the light ray is the most distorted by gravitating bodies is of a
distance $\sim n\sigma=\sqrt{\beta/2}$, which then gives an approximate
relativistic $O(1/c^{2})$ correction 
\begin{equation}
\delta\alpha_{g}\simeq\frac{{3\kappa n\sigma^{2}(\theta^{2}+2\theta\delta\alpha_{c})}}{{c^{2}}}\Big(\frac{{\Sigma}}{{\Sigma_{c}}}\Big)^{2},\label{eq: 5.64}
\end{equation}
where $\delta\theta\simeq\delta\alpha_{c}$. An approximate Lyapunov
exponent is then 
\begin{equation}
\lambda\simeq\lim_{n~\rightarrow~\infty}\frac{1}{n}log\Big(1+\kappa\big(\frac{\Sigma_{c}}{{\theta\Sigma_{c}}}\big)\Big[e^{\xi^{2}log(n)\beta}+3\kappa n\sigma^{2}(\theta^{2}+2\theta e^{\xi^{2}log(n)\beta})\Big(\frac{{\Sigma}}{{\Sigma_{c}}}\Big)^{2}\Big]\Big).\label{eq: 5.65}
\end{equation}
Here there is an amplification of the ray trace uncertainty, or chaos,
by the introduction of $O(1/c^{2})$ term as seen in the term $2\theta e^{\xi^{2}log(n)\beta}(\Sigma/\Sigma_{c})^{2}$.
For $\delta\alpha_{c}\simeq\theta$ the contributions to the chaotic
ray traced path from relativistic corrections and chaos are comparable
and will contribute equally to the randomness of the caustic gravitational
lens. 

The difference in the perturbed aperture vectors $\Delta\delta\vec{\xi}=\delta\vec{\xi}_{i}-\delta\vec{\xi}_{s}\nabla_{\delta\xi}\Phi$
determines the magnification $M=d(\delta\xi_{i})/d(\delta\xi_{s})$.
From Hamilton's equations this is generalized to 
\begin{equation}
\Delta\delta\vec{\xi}=\nabla_{\xi}H\simeq2\kappa\xi log(n)d_{l}^{2}\alpha_{c}\Big(1+\frac{{3nd_{l}^{2}\theta}}{{c^{2}}}\Big(\frac{{\Sigma}}{{\sigma_{c}}}\Big)^{2}\Big)\label{eq: 5.66}
\end{equation}
with the deviation magnification computed accordingly. For this written
according to the radius of curvature $R$ of a surface for a ray curve
along a line of sight we have that 
\begin{equation}
\frac{{2\pi A\sqrt{2|R|}}}{{\Sigma_{c}}}\Theta(\delta\vec{\xi}_{i})=\frac{{3nd_{l}^{2}\theta}}{{c^{2}}}\Big(\frac{{\Sigma}}{{\Sigma_{c}}}\Big)^{2}\label{eq: 5.67}
\end{equation}
The magnification for $\eta=2\pi\sqrt{2|R|}/(\Sigma_{c})$ is ${\cal M}=1+\eta\Theta+O(\eta^{2})$.
The curvature $R$ defines a tangent for the ray trace, which defines
a caustic when the line of sight is along this tangent. The caustics
along lines of site occur at swallowtail folds in the magnification
map.

For the orbits in different planes the above must be generalized some.
Similarly the angular components are given by the tangent vector parallel
to $p_{i}$ and from there we may find the vector to each dark matter
clump in its plane of motion with coordinates $\{u,\theta\}$ and
$\{v,\theta'\}.$ Further, as the angular momentum vector is given
by ${\vec{l}}={\vec{r}}_{1}\wedge{\vec{p}}_{1}/m$ (similar for ${\vec{L}}$),
then $\vec{L}$ is rotated relative to $\vec{l}$ by the Euler angles
$\alpha$, $\beta$ and $\gamma$. The rotation matrix is then 
\begin{equation}
{\bf [R]}=\left[\begin{matrix}\cos(\gamma)\end{matrix}\right]\left[\begin{matrix}1\end{matrix}\right]\left[\begin{matrix}cos(\alpha)\end{matrix}\right].\label{eq: 5.68}
\end{equation}
With these we may be able to put the problem in a general setting.
This part is yet to be worked, and their may be resources to aid in
this effort. 

The formation of filaments and domain walls of galaxies is then proposed
to occur by this mechanism. The massive gravity waves in the very
early universe, such as in the post inflationary period, deviate the
motion of relativistic particles in a manner similar to the optical
focusing of light. These focal points of matter then set up their
own gravity fields which persist through the subsequent expansion
of the universe. A mesh of caustics with swallowtail cusps heuristically
may be seen to produce a web of regions where mass-energy is concentrated.
The distribution of dark matter may then be established by caustics
of gravitons and gravity waves in the early universe.

\section{Conclusion remarks}

In this paper the production of massive relic coherent gravitons in
in a particular class of $f(R)$ gravity which arises from string
theory and their possible imprint in CMB have been discussed. The
key point is that in the very early universe these relic gravitons
could have acted as slow gravity waves. They may have then acted to
focus the geodesics of radiation and matter. Therefore, their imprint
on the later evolution of the universe could appear as filaments and
domain wall in the Universe today. In that case, the effect on CMB
should be analogous to the effect of water waves, which, in focusing
light, create optical caustics which are commonly seen on the bottom
of swimming pools. This issue has been carefully analyzed by showing
gravitational lensing by relic GWs, i.e. how relic massive GWs perturb
the trajectories of CMB photons.

The consequence of the type of physics discussed has been outlined
from the point of view of an amplification of what could be called
optical chaos. 

For the sake of completeness, we stress that multiple imaging by gravitational
waves and the associated caustic structure have been studied other
authors in frameworks different with respect to the approach of this
paper, see for example \cite{key-43} and references within.

\section{Acknowledgements}

C. Corda is partially supported by a Research Grant of The R. M. Santilli
Foundation Number RMS-TH-5735A2310. The authors thank the second and
the third reviewer for important criticisms and suggestions which
permitted to improve this paper.


\begin{thebibliography}{10}
\bibitem{key-1}D. Lust and S. Theisen, Lect. Notes Phys. 346, 1 (1989).

\bibitem{key-2}\foreignlanguage{italian}{M. Green, J. Schwarz and
E. Witten, ``\emph{Superstring Theory}'', Volume 1, Cambridge U.
Press, Cambridge (1987).}

\bibitem{key-3}S. Nojiri, S. D. Odintsov, Phys. Rep. 505, 59 (2011).

\bibitem{key-4}A. Einstein,\textit{ }\textit{\emph{Preuss. Akad.
Wiss. Berlin, Sitzber., 778 (1915).}}

\bibitem{key-5}\foreignlanguage{italian}{H. Nicolai, G. F. R. Ellis,
A. Ashtekar and others, \emph{``Special Issue on quantum gravity}'',
Gen. Rel. Grav. 41, 4, 673 (2009).}

\selectlanguage{italian}%
\bibitem{key-6}L. H. Ryder, \emph{Quantum field theory, }Cambridge
University Press (1996).

\bibitem{key-7}\foreignlanguage{english}{C. Corda, Int. Journ. Mod.
Phys. D, 18, 14, 2275 (2009).}

\selectlanguage{english}%
\bibitem{key-8}P. J. E. Peebles and B. Ratra, Rev. Mod. Phys. 75,
559 (2003).

\bibitem{key-9}A. Einstein, Sitzungsber. Preuss. Akad. Wiss., 142,
235 (1931).

\bibitem{key-10}P. Jordan, Naturwiss. 26, 417 (1938).

\bibitem{key-11}M. Fierz, Helv. Phys. Acta 29, 128 (1956).

\bibitem{key-12}C. Brans and R. H. Dicke, Phys. Rev. 124, 925 (1961).

\bibitem{key-13}L. Landau and E. Lifsits\textit{, Classical Theory
of Fields} (3rd ed.), London: Pergamon (1971).

\bibitem{key-14}J. J. Ferrando, J. A. Sáez, Gen. Rel. Grav. 42, 1469(2010). 

\bibitem[15]{key-15}S. Capozziello, C. Corda, M. F. De Laurentis,
Phys. Lett. B 669, 255 (2008). 

\bibitem[16]{key-16}C. Corda, J. Cosmol. Astropart. Phys. JCAP04009
(2007).

\bibitem[17]{key-17}\foreignlanguage{italian}{C. W. Misner , K. S.
Thorne, J. A. Wheeler, \textit{``Gravitation''}, Feeman and Company
(1973).}

\bibitem[18]{key-18}B. Allen, in ``\emph{Proceedings of the Les
Houches School on Astrophysical Sources of Gravitational Waves}'',
p.373, eds. Jean-Alain Marck and Jean-Pierre Lasota, Cambridge University
Press (1998).

\selectlanguage{italian}%
\bibitem[19]{key-19}D. H. Lyth and A. R. Liddle, \emph{``Primordial
Density Perturbation}'', Cambridge University Press (2009).

\selectlanguage{english}%
\bibitem[20]{key-20}\foreignlanguage{italian}{C. Corda, Eur. Phys.
J. C 65, 257 (2010). }

\bibitem[21]{key-21}\foreignlanguage{italian}{Planck Collaboration,
P. A. R. Ade, et al., ArXiv e-prints (2013), arXiv:1303.5062. }

\bibitem[22]{key-22}\foreignlanguage{italian}{Planck Collaboration,
P. A. R. Ade, et al., ArXiv e-prints (2013), arXiv:1303.5072. }

\bibitem[23]{key-23}\foreignlanguage{italian}{A. Jannussis and E.
Skuras, Il Nuovo Cimento B, 95, 63 (1986). }

\bibitem[24]{key-24}\foreignlanguage{italian}{S. J. Chang, ``\emph{Introduction
to quantum field theory}'', World Scientific (1990).}

\selectlanguage{italian}%
\bibitem[25]{key-25}C. F. Lo, Quantum Opt. 3, 333 (1991).

\bibitem[26]{key-26}J. Katriel, Phys. Lett. A, 307, 1 (2003). 

\bibitem[27]{key-27}C. Gerry, and P. Knight, ``\emph{Introductory
Quantum Optics}'', Cambridge University Press (2005).

\bibitem[28]{key-28}K. Wilson, Phys. Rev. D 10, 2445 (1974).

\bibitem[29]{key-29}O. Darrigol, ``\emph{Electrodynamics from Ampere
to Einstein}'', Oxford (2000).

\bibitem[30]{key-30}P. Schneider, J. Elhers, E. E. Falco\textit{,
Gravitational Lenses }, Springer-Verlag, Berlin (1992).

\bibitem[31]{key-31}C. Corda, H. Mosquera Cuesta, R. Lorduy-Gòmez,
Astropart. Phys. 35, 362-370 (2012).

\bibitem[32]{key-32}J. E. Greivenkamp\emph{, Field Guide to Geometrical
Optics}. SPIE Field Guides vol. FG01. SPIE. pp. 19\textendash{}20.
ISBN 0-8194-5294-7 (2004).

\bibitem[33]{key-33}I. I. Shapiro\emph{,} Phys. Rev. Lett. \textbf{13},
789\textendash{}791 (1964).

\bibitem[34]{key-34}\foreignlanguage{english}{M. Cencini et al.,
\emph{Chaos From Simple models to complex systems, }World Scientific.
ed. (2010). }

\bibitem[35]{key-35}\foreignlanguage{english}{A. Einstein, Science
84 (2188), 506\textendash{}7 (1936).}

\bibitem[36]{key-36}\foreignlanguage{english}{V. Berezinsky, V. Dokuchaev,
Y. Eroshenko, Phys. Rev. D 77, 083519 (2008).}

\bibitem[37]{key-37}\foreignlanguage{english}{J.A. Muñoz, E. Mediavilla,
C. S. Kochanek, E. Falco, A. M. Mosquera, arXiv:1107.5932 (2011).}

\bibitem[38]{key-38}V. I. Arnold, \emph{Catastrophe Theory}, 3rd
ed. Berlin: Springer-Verlag (1992).

\bibitem[39]{key-39}R. A Cabanac et al., Astron. Astrophys. 436 (2),
L21\textendash{}L25 (2005).

\selectlanguage{english}%
\bibitem[40]{key-40}\foreignlanguage{italian}{B. Altieri et al.,
Astron. Astrophys. 518, L17 (2010).}

\bibitem[41]{key-41}\foreignlanguage{italian}{S. G. Brush, Rev. Mod.
Phys. 39, 883 (1967).}

\selectlanguage{italian}%
\bibitem[42]{key-42}C. Corda, Mod. Phys. Lett. A 22, 16, 1167-1173
(2007).

\bibitem[43]{key-43}\foreignlanguage{english}{A. I. Harte, Class.
Quant. Grav. 30, 075011 (2013).}

\selectlanguage{english}%
\bibitem[44]{key-44}Private communication with the second referee.

\bibitem[45]{key-45}K. Stelle, Phys. Rev. D 16, 953 (1977).

\bibitem[46]{key-46}K. Stelle, Gen. Rel. Grav. 9, 353 (1978). 

\bibitem[47]{key-47}H. J. Schmidt, Phys. Rev. D 83, 083513 (2011). 

\bibitem[48]{key-48}P. X . Jiang, J. W. Hu, Z. K. Guo, Phys. Rev.
D 88, 123508 (2013).

\bibitem[49]{key-49}K. Nozari, N. Rashidi, Phys. Rev. D 88, 084040
(2013).

\bibitem[50]{key-50}M. Sharif, G. Abbas, Eur. Phys. J. Plus 128,
102 (2013).

\bibitem[51]{key-51}K. Nozari, F. Kiani, N. Rashidi, Adv. High Energy
Phys. 968016, 12 (2013).

\bibitem[52]{key-52}W. Yao, J. Jing, JHEP 05, 101 (2013).

\bibitem[53]{key-53}X. X. Zeng, W. B. Liu, Physics Letters B 726,
481 (2013).

\bibitem[54]{key-54}L.Dabrowski, A. Sitarz, J. Math. Phys. 54, 013518
(2013).

\bibitem[55]{key-55}C. Simeone, Phys. Rev. D 83, 087503 (2011).

\bibitem[56]{key-56}S. H. Hendi, Phys. Lett. B 677, 123 (2009).

\bibitem[57]{key-57}S. H. Hendi, B. E. Panah, Phys. Lett. B 684,
77 (2010).

\bibitem[58]{key-58}G. F. R. Ellis, J. Murugan, C. G. Tsagas, Class.
Quant. Grav. \textbf{21}, 233-250 (2004).

\bibitem[59]{key-59}S. Nojiri and S. D. Odintsov,\foreignlanguage{italian}{
Phys. Rev. D \textbf{68}, 123512 (2003).}

\bibitem[60]{key-60}\foreignlanguage{italian}{C. Corda, Gen. Rel.
Grav. \textbf{40}, 10, 2201-2212 (2008).}

\selectlanguage{italian}%
\bibitem[61]{key-61}S. Nojiri, S.D. Odintsov, ECONF C0602061, 06,
(2006); Int. J. Geom. Meth. Mod. Phys. 4, 115 (2007).

\bibitem[62]{key-62}\foreignlanguage{english}{Private communication
with the third referee.}

\selectlanguage{english}%
\bibitem[63]{key-63}X. Wu, T. Y. Huang, Phys. Lett. A 313, 77-81
(2003).

\bibitem[64]{key-64}S. Capozziello, M. De Laurentis, S. Nojiri, S.
D. Odintsov, Gen. Rel. Grav. 41, 2313 (2009).\end{thebibliography}
\end{document}